%% file: ms.tex
\documentclass[preprint,11pt]{aastex}
\shorttitle{Complex Envelopes}
\shortauthors{Tobin et al.}

\begin{document}

\title{Complex Structure in Class 0 Protostellar Envelopes\footnotemark[1]}
\author{John J. Tobin\altaffilmark{2}, Lee Hartmann\altaffilmark{2}, Leslie W. Looney\altaffilmark{3}, Hsin-Fang Chiang\altaffilmark{3}}

\begin{abstract}
We use archived IRAC images from the \textit{Spitzer Space
Telescope} to show that many Class 0 protostars exhibit
complex, irregular, and non-axisymmetric structure within their
dusty envelopes.  Our 8 $\mu$m extinction maps probe some
of the densest regions in these protostellar envelopes.
Many of the systems are observed to have highly irregular and non-axisymmetric
morphologies on scales $\ga$ 1000 AU, with a quarter of the
sample exhibiting filamentary or flattened dense structures. 
Complex envelope structure is observed in regions spatially distinct
from outflow cavities, and the densest structures often show
no systematic alignment perpendicular to the cavities. These results indicate
that mass ejection is not responsible for much of the irregular morphologies
we detect; rather, we suggest that the observed envelope complexity
is mostly the result of collapse from protostellar
cores with initially non-equilibrium structures. 
The striking non-axisymmetry in many envelopes could provide favorable
conditions for the formation of binary systems.  We also note
that protostars in the sample appear to be formed preferentially
near the edges of clouds or bends in filaments, suggesting formation
by gravitational focusing.
\end{abstract}

\footnotetext[1]{This paper includes data gathered with the 6.5 meter Magellan Telescopes located at Las Campanas Observatory, Chile}
\altaffiltext{2}{Department of Astronomy, University of Michigan, Ann
Arbor, MI 48109; jjtobin@umich.edu}
\altaffiltext{3}{Department of Astronomy, University of Illinois at
Champaign/Urbana, Urbana, IL 61801 }

\section{Introduction}

Sphericity and axisymmetry have been standard assumptions on which
our theoretical understanding of star formation has rested for some time.
One of the early models of protostellar
collapse by \citet{shu1977} was based on the singular isothermal sphere
developed and extended to include rotation by \citet[][TSC]{tsc1984}.
Further modifications have been introduced over time, including
models with flattened envelopes or `pseudo-disks' as in \citet{galli1993} and
\citet{hartmann1996} while still assuming axisymmetry.
Spherical/axisymmetric envelope models have been used extensively
to calculate spectral energy distributions (SEDs) of embedded
protostars or disks \citep[e.g.][]{kch1993,whitney1993,whitney2003}.
In particular, the TSC envelope model has been highly successful
in modeling the SEDs of Class 0 and Class I
protostars; by including outflow cavities, such models are
also able to reproduce near and mid-infrared scattered light images
\citep[e.g.][]{adams1987,furlan2008, kch1993,stark2006,tobin2007,tobin2008}.
However, it is not clear whether or not envelopes around protostars
are accurately described by symmetric models.

Recently, observations with the \textit{Spitzer Space Telescope} have given a high-resolution
view of envelope structure around two Class 0 protostars in extinction at 8$\mu$m against Galactic
background emission, L1157 appears flattened and L1527 has an 
asymmetric distribution of material \citep{looney2007, tobin2008}.
This method enables us to observe the structure of collapsing protostellar envelopes
on scales from $\sim$1000AU to 0.1 pc for the first time with a mass-weighted tracer. In contrast,
single-dish studies of envelopes using dust emission in the sub/millimeter
regime generally have lower spatial resolution. The continuum
emission depends upon temperature as well as mass, while
molecular tracers are affected by complex chemistry.
Interferometry can provide higher resolution of both continuum and molecular 
tracers, but large scale structure is resolved out, in contrast to the
8 $\mu$m extinction maps.

In this paper, we analyze archival IRAC images of 22 Class 0 protostars whose dusty envelopes
can be detected in extinction at 8$\mu$m.
Most of the envelopes in our sample are found to be irregular and non-axisymmetric.
We demonstrate that the extinction we observe is indeed due to
the circumstellar envelope
and not background fluctuations by comparing near-IR extinction
measurements with those at 8$\mu$m.  Near-infrared imaging
of lower-extinction regions is used to correct
for foreground emission and/or instrumental effects.
 We also derive quantitative measures of the
envelope asymmetries using projected moment of inertia ratios.
Our results indicate that infalling envelopes are frequently complex and
non-axisymmetric, which might be the result of gravitational collapse
from complex initial cloud morphologies. We also suggest that protostars
exhibit a preference to form near the edges of clouds or bends in
filaments, which could be due to the effects of gravitational focusing.

\section{Observations and Data Reduction}

The primary dataset used in this study is comprised of archived \textit{Spitzer Space Telescope}
8$\mu$m images taken with the IRAC instrument. We have also taken near-IR (H \& Ks-band) images of the protostars and surrounding regions
with a rich stellar background as an additional constraint on column density using near-IR extinction mapping
via the near-infrared color excess (NICE) method \citep{lada1994}.

\subsection{\textit{Spitzer} IRAC Observations}

Motivated by prior detections of envelopes in extinction, we downloaded the pipeline-reduced data
for cataloged Class 0 protostars \citep[e.g.][]{froebrich2005, seale2008} as well as the c2d \citep{evans2003}
observations of dense cores and molecular clouds to determine if
a protostar has an 8$\mu$m extinction envelope associated with it.
Of the cataloged protostars, we have clearly detected 22 envelopes in extinction
within the nearby star forming clouds (Taurus, Perseus, Cepheus, Chameleon, Orion).
We were not able to obtain meaningful results for sources in which the background
emission is too faint to reliably derive $8 \mu$m extinction, or in cases where the
protostar is too bright, thus swamping its envelope structure with emission on the wings
of the point-spread function (PSF), or in very crowded regions with many
protostars, outflows, and extended foreground (PAH) emission.
Pre-stellar/starless cores (e.g., \citet{stutz2009} and \citet{bacmann2000})
are not considered in this study.

For each source with identified extinction, we downloaded the basic calibrated data (BCDs) and mosaicked the 
individual frames using MOPEX 
after running the \textit{Spitzer} IRAC artifact mitigation tool written by S. Carey. Because the IRAC data use sky
darks rather than true darks, all dark frames contain some level of zodiacal light emission that
is subtracted from the BCDs. For our purposes, it is necessary to eliminate
the zodiacal light from our images. Thus, during the artifact mitigation process we subtracted the difference
between the zodiacal background of the target and the sky dark zodiacal background as written in the BCD headers
which are determined from a zodiacal light model \citep{meadows2004}. 

We list the selected sources, observations dates, integration times, and AOR keys in Table 1. 
The sources in which the data were originally observed by the authors are also 
identified with the program number.  Also, some objects had multiple epochs of observation. In these cases,
the datasets were combined if the
data were taken close enough in time such that the difference between estimated zodiacal
emission was negligible. If the observations
were taken more than a few days apart, we used the set of observations with the longest integration time.

\subsection{Near-IR Observations}

To complement our \textit{Spitzer} 8$\mu$m data, we observed selected protostars from our sample in H and Ks-bands
for the purpose of measuring extinction toward background stars viewed through the envelope. 
We have identified the protostars for which near-IR data were taken in Table 1. Data for L1152, L1157, L1165, L723, and L483
were taken at the MDM Observatory on Kitt Peak using the near-IR instrument TIFKAM on the the 2.4m Hiltner telescope during
photometric conditions between 29 May 2009 and 4 June 2009. TIFKAM
provides several imaging modes, we used the F/5 camera mode which provides a $\sim$5$^{\prime}$ field-of-view (FOV) over the
1024$^2$ array. We observed the targets in a 5-point box dither pattern with 30$^{\prime\prime}$ steps, taking 5x30 second coadded
images at each point in H and Ks-bands. Total integration times were generally 50 minutes in Ks-band and 40 minutes in H-band,
these were varied depending on seeing and sky-background.  We were not concerned with preserving extended emission from the protostars, therefore we median combined
the images of a single dither pattern to create a sky image for subtraction. The data were reduced using the UPSQIID package
in IRAF\footnote{IRAF is distributed by the National
Optical Astronomy Observatories,
which are operated by the Association of Universities for Research
in Astronomy, Inc., under cooperative agreement with the National
Science Foundation.}.

The observations of BHR71, IRAS 09449-5052, and HH108 were taken with the ISPI camera \citep{vdbliek2004} at CTIO using the 4m Blanco
Telescope during photometric conditions on 11 June 2009. The ISPI camera features a 10.5$^{\prime}$ FOV on a 2048$^2$ array. 
We observed the protostars using a 10-point box dither pattern with 60$^{\prime\prime}$ steps with 3x20 second coadded images 
in Ks-bands and generally 2x30 second coadded images in H-band. The total integration time
for BHR71 was 30 minutes in each band, and 20 minutes for IRAS 09449-5052. Again, we median-combined the on-source frames to 
create the sky image; however, due to the larger field and steps in the dither pattern extended emission is preserved in 
these data. We used standard IRAF tasks for flat-fielding and sky subtraction. We could
not simply combine the data using an alignment star due to optical distortion.
To correct for this, we fit the world-coordinate system (WCS)
to each flat-fielded, sky-subtracted frame using wcstools \citep{mink1999} and the 2MASS catalog. 
Then we used the IRAF task CCMAP to fit a 4th order polynomial to the coordinate 
system. A difficulty encountered was the lack of 2MASS stars in the center of the images since our targets are protostars with 
highly opaque envelopes. In addition, the 2MASS catalog tends to have some false source identifications associated with diffuse scattered
light in the outflow cavity of protostars. Thus, since not all polynomial fits were acceptable, we applied the best fitting solution 
to all images. We then used the stand-alone program SWARP \citep{bertin2002} to combine the individual frames while accounting for 
the distortion. 

Lastly, we conducted additional observations of BHR71 with the PANIC camera on the 6.5m Magellan (Baade) telescopes.
The data were taken during photometric conditions on 17 and 18 January 2009. The PANIC instrument
has only a $\sim$2$^{\prime}$ FOV, thus we took 3 fields of BHR71, one centered on the protostar, one 2$^{\prime}$
east and 45$^{\prime\prime}$ south, and a last field  2$^{\prime}$
west and 45$^{\prime\prime}$ north. The data were taken in a 9-point dither
pattern with 2x20 second images taken at each position (coadds are not supported with PANIC).
The total integration time for each
field was 12 minutes in H and Ks-bands. The seeing during these observations was $\sim 0.4^{\prime\prime}$; thus,
these images detect fainter stars despite shorter integration times than to the ISPI observations.
Separate offset sky observations were taken for the 
central field, the east and west fields were median combined to create the sky image. The data were reduced using the UPSQIID package
in IRAF.

For all the above datasets, photometry of stars in the images was measured using the DAOPHOT package in IRAF. We used DAOPHOT to
identify point-sources, create a model PSF from each combined image, and measure instrumental magnitudes using PSF photometry.
The magnitude zeropoints were determined by matching the catalog from DAOPHOT to the 2MASS catalog and fitting a
Gaussian to a histogram of zeropoint measurements. The H and Ks-band catalogs were then matched using a custom IDL program
which iteratively finds corresponding star in each catalog and computes the $H - Ks$ color.
Only sources with detections at H and Ks-bands were included in the final catalog.

\section{Results}

In Figures 1-5, we display the 22 systems for which we have detected
an envelope in extinction.  
The 3.6$\mu$m image is shown in the left panels, the 8.0$\mu$m image 
with optical depth contours (see \S4) from 8$\mu$m data in the middle panel,
the 8.0$\mu$m image with 
SCUBA 850$\mu$m contours overlaid in the right panel. In the case of HH108, (Figure 1)
there were no IRAC data, so we plot the Ks-band image, MIPS 24$\mu$m image and optical depth
contours and  the Ks-band image with SCUBA 850$\mu$m contours overlaid.
The 3.6$\mu$m/Ks-band image for each source shows the 
scattered light cavity and, for some of the data with very deep integrations, the envelopes
are outlined by diffuse scattered light. 
The 8.0$\mu$m images and optical depth contours show
the envelope structure in extinction, and the overlaid SCUBA  data from \citet{difrancesco2008}
show how the thermal envelope emission correlates with the 8.0$\mu$m extinction.

The most striking feature of the 8$\mu$m extinction maps is the irregularity of 
envelopes in the sample. Most envelopes show high degrees of non-axisymmetry;
in most cases, spheroids would not provide an adequate representation of the
structure. Some of the most extreme examples have most extincting material mostly on
one side of the protostar (e.g. CB230, HH270 VLA1) or the densest structures are
curved near the protostar (e.g. BHR71, L723). The structures seen in extinction at 8$\mu$m
do not seem to be greatly influenced by the outflow. The 3.6$\mu$m images show that the
outflow cavities of these sources are generally quite narrow, with
a relatively small evacuated region; the dense
material detected in extinction is often far from the outflow cavities and thus
seems unlikely to be produced by outflow effects (see \S 6.1 for further discussion).

For convenience, we categorize the systems into 5 groups according to their morphology, 
though some systems have characteristics of multiple groups.
Figure 1 shows the envelopes that have a highly filamentary or flattened morphology; Figure
2 shows envelopes that have most material (in projection)
on one side of the protostar; Figure 3 shows the protostars
whose envelopes are more or less spheroidal in projection;
Figure 4 shows the protostars which appear to be binary (i.e., one protostar
is present and about 0.1 pc away there is an extinction 
peak probably corresponding to a starless core).
Lastly, Figure 5 shows envelopes that do not strictly fall within the above categories,
and are simply classified as irregular.

This dense complex structure is less apparent in the SCUBA maps primarily
because it has a resolution of $\sim$ 12$^{\prime\prime}$
at 850$\mu$m while IRAC has a diffraction limited resolution of $\sim$2$^{\prime\prime}$. SCUBA also has limited
sensitivity to extended structure due to the observation method;
 thus IRAC can better detect non-axisymmetries on smaller and larger scales than 
seen in SCUBA maps. In addition,  the strongest emission from most envelopes appears
to be axisymmetric because the protostar is warming the envelope \citep{chiang2008,chiang2009};
the extinction maps are not affected by the envelope temperature distribution.

\section {Optical Depth Maps}

Though the morphology of extincting material is clear from direct inspection of the images,
it is desirable to convert the 8 $\mu$m  intensities to optical depth maps
to make quantitative measurements. Initially we assumed
that the observed intensities can be interpreted as pure extinction with
no source emission, i.e.
\begin{equation}
\label{taucalc}
\frac{I_{obs}}{I_{bg}}=e^{-\tau}
\end{equation}
where  $I_{obs}$ is the observed intensity per pixel 
and $I_{bg}$ is the measured background intensity,
both corrected for the estimated zodiacal light intensity as discussed in \S2.1.
However, our attempts to model the filamentary structure in L1157 
(\S 5.2) and the very low column densities measured led us to conclude that our maps contain 
more foreground emission and/or zero-point correction than we originally
thought. This may not be surprising, as 
measurements of the zodiacal light intensity during the 
\textit{Spitzer} First Look Survey were 36\%
higher than predicted by the model \citep{meadows2004}. 
Either there is residual zodiacal light not accounted for by the model, or there is
foreground dust emission, scattered light within the detector
material \citep{reach2005}, or some combination of these factors.
Unfortunately, because IRAC operates without a shutter, it is impossible to determine
the true level of diffuse emission. 

Thus, in our initial analysis we were calculating
\begin{equation}
\frac{I_{obs} + I_{fg}}{I_{bg} + I_{fg}}=e^{-\tau}
\end{equation}
which will yield an erroneously low value of $\tau$.
This is because as $I_{fg}$ increases, the ratio of 
$I_{obs}^{\prime}$/$I_{bg}^{\prime}$
increases, causing the measured optical depth to decrease.

We therefore obtained the ground-based near-infrared imaging data to 
develop an independent estimate of the extinction in lower-column density
areas, and thus made a better estimate of the foreground contribution.
The Ks-band images of BHR71 and L483 with optical depth contours overlaid in Figure \ref{nirimages}
show that many background stars can be detected through to dense envelope. 
The details of the foreground correction are discussed in the Appendix.

With the estimated foreground contribution was subtracted, we were able to calculate 
a more accurate optical depth map following Equation \ref{taucalc}.
We now must calculate the background emission in the image; fortunately in most images 
the background is relatively uniform. We estimated the background emission 
by fitting a Gaussian to a pixel histogram constructed 
from an area in the image not affected by the extinction of the envelope.
For the sources IRAS 05295+1247, HH211, and L1165, 
the background has clear gradients.
To account for this, we constructed a background model by performing a two-pass
median filter of the entire image with a convolving beam  90$^{\prime\prime}$ (IRAS 05295+1247)
and 144$^{\prime\prime}$ (L1165, HH211) in size; this method 
is similar to that of \citet{simon2006, ragan2009, butler2009}. 
We used the first pass to identify stars
in our background measurement field, then rejected those pixels in the
second pass. The median background computed from the first and second passes are generally within
a few percent of each other. The large convolving
beam ensures that we filter out the envelope from the background model. 

We then divided each pixel by the background value,
or, in the case of IRAS 05295+1247/L1165, we divided the intensity
map by the background model.  
Then taking the natural logarithm of each pixel intensity
yielded the optical depth map.
In the case of HH108, 
we constructed an optical depth map from the MIPS 24$\mu$m image because there are no
IRAC data for this object.

We give the values of $I_{fg}$, $I_{bg}^{\prime}$ (see Appendix), $\tau_{max}$, and $\sigma\tau$ for each source in Table 2.
In all cases, $I_{fg}$ is comparable to measured intensity at 
the darkest spot in the uncorrected 8$\mu$m image. Thus,
we can infer than the darkest part of an envelope is completely
opaque. To set an upper limit on the optical depth in these areas, we
use the pixel value in the uncertainty image as $I_{obs}$ and 
compute the optical depth; this is the maximum optical depth ($\tau_{max}$)in an image.
To correct our images without near-IR data, we use the result that in the most
opaque areas of the images $I_{fg}$ is the observed intensity and take this
value as an estimate of the foreground contribution.

\section{Quantitative Results}

\subsection{Non-Axisymmetric Structure}

To quantify the evident envelope asymmetry of many sources, 
we calculate projected moment of inertia ratios using
$\tau$ as a surrogate for mass. We calculate the ratios by computing
\begin{equation}
\frac{I_{\perp}}{I_{\parallel}}=\frac{\sum_{i}^{} \tau_i(x_i-x_{\circ})^2}{\sum_{j}^{} \tau_j(y_j-y_{\circ})^2}
\end{equation}
where $x_{\circ}$ and $y_{\circ}$ are the coordinates 
of the protostar and $I_{\perp}$ is the moment of
inertia of material distributed perpendicular the outflow along the abscissa and $I_{\parallel}$
is for material located parallel to the outflow axis along the ordinate axis. The subscripts $i$ and $j$ 
denote the independent points where the optical depth is measured. We rotate the optical depth images
such that the outflow is along the ordinate axis of the image to simplify interpretation.
We also calculate the moment of
inertia ratios for $I_{\perp}$ left and right  of the protostar ($I_{\perp,l}$/$I_{\perp,r}$) and the same
 for $I_{\parallel}$ ($I_{\parallel,l}$/$I_{\parallel,r}$). 
We measure the ratios out to a radius of 0.05 pc for most protostars at
the adopted distance in Table 1 and we set an inner radius of 0.01 pc.The outer radius is restricted so
the moments of inertia are sensitive to the densest structures closest to the protostar and not influenced
by extended diffuse material. Also, areas where emission is present are masked by requiring that the optical depths be positive.

 Each ratio quantifies
a different aspect of the distribution of material around the protostar. $I_{\perp}$/$I_{\parallel}$
describes how much material is located along the outflow axis versus perpendicular to it, a ratio greater than
1 would correspond to more material away from the protostar, perpendicular to the outflow. A ratio less than 1 
corresponds to having more material close to the outflow axis, extended in the direction of the outflow.
$I_{\perp,l}$/$I_{\perp,r}$ describes
the asymmetry about the outflow axis by comparing the measurements on the right and left sides of the protostar
and $I_{\parallel,l}$/$I_{\parallel,r}$ describes the asymmetry 
about the outflow but in the vertical direction. Taken together these ratios describe the distribution
of material around the protostar, convenient for comparison to theoretical models.

The projected moment of inertia ratios for the envelopes are given in Table 3. 
There are more envelopes that are extended perpendicular to the outflow than along it
(most $I_{\perp}$/$I_{\parallel}$ ratios are $> 1$). 
Many objects exhibit large ratios and they
can be described as highly flattened/elongated. HH270 VLA1 and L1152 seem to be the only examples of 
envelopes strongly extended in the direction of their outflows in our sample. However, both
L673-SMM2 and Serp-MMS3 have components of their envelopes oriented parallel and perpendicular
to their outflows. This yields a $I_{\perp}$/$I_{\parallel}$ ratio $\sim$1 but in the other ratios,
 non-axisymmetry is evident.

The most symmetric envelope is IRAS 16253-2429,
as shown by its moment of inertia ratios all being near 1. 
The two most non-axisymmetric envelopes appear to be HH270 VLA1 and CB230.
HH270 VLA1 has most of its material located southwest of the protostar and CB230
has most of its material located to the west and in a moderately-flattened configuration.

\subsection{Flattened Structure}
There are six sources that have remarkably flat structure compared to the rest of the sample, shown in Figure 1: L1157, L723, HH108,
Serp-MMS3, L673-SMM2, and BHR71. With respect their molecular outflows, L1157 and BHR71 are viewed nearly edge-on; the primary protostar
in HH108 seems to be edge-on as well. Serp-MMS3 and L673-SMM2 are not well studied;
however, 3.6$\mu$m image of Serp-MMS3 indicates that it is at least inclined by 60$^{\circ}$ or more and L673-SMM2 harbors several 
protostars and their inclinations are not known. The orientation of L723 is uncertain as there are two embedded sources driving outflows
and due to the complexity of the data we omit this object from our analysis.

For these sources, we ask the question:
are these envelopes flattened sheets/pseudo-disks \citep{hartmann1994,hartmann1996,galli1993} or filaments? To attempt 
to answer this question, 
we compare the observed vertical structure of the flattened envelopes 
to analytic prescriptions for isothermal hydrostatic filaments and sheets.

The scale height of an isothermal filament in hydrostatic equilibrium is 
\begin{equation}
H_f=\frac{c_s^2}{2 G \Sigma_{0,f}}
\end{equation}
where $c_s$ is the isothermal sound speed, we assumed T=10K, and $\Sigma_{0,f}$ 
is the peak surface density measured at the center of the filament \citep{hartmann2008}.
When parametrized in terms of 8$\mu$m optical depth and 
assuming $\kappa_8$ = 10.96 cm$^2$g$^{-1}$, 
\begin{equation}
H_f=\left(\frac{0.96}{\tau_{0,f}}\right)\left(\frac{T}{10 K}\right)^{-1} \times 0.01 pc.
\end{equation}
Thus, a 10 K filament with a scale height of 0.01 pc will have $\tau$ $\sim$1 at 8$\mu$m.

Similarly, the scale height of an isothermal infinite sheet is given by
\begin{equation}
H_s=\frac{c_s^2}{\pi G \Sigma_{0,s}}.
\end{equation}
In this case, $\Sigma_{0,s}$ is \textit{not} the surface density measured at the center of the sheet viewed edge-on
but it is the surface density through the z-direction of the sheet. Thus, we must make
an assumption about depth of the sheet into the line of sight. Making the
same assumption about temperature as the filament case, we can
write the scale height of a sheet as
\begin{equation}
H_s\sim\frac{1.2}{\tau_{0,f}}\left ( \frac{d}{t}\right)(0.01 pc)
\end{equation} 
where $d$ is the line-of-sight depth through the sheet and $t$ is the thickness of
the sheet in the plane of the sky. Together,
$d$ and $t$ specify the aspect ratio of the sheet; $d$ is assumed to be 0.1 pc
(the diameter of most envelopes in the sample) and $t$ is taken to be the FWHM 
of the Gaussian fit to the vertical structure described in the next paragraph.
For an aspect ratio of 10, the peak optical depth would have to be $\sim$12
in order to have a scale height of 0.01 pc.

We analyzed the structure by averaging the extinction map along the
extended dimension in three pixel bins, and then fitting a Gaussian
to the perpendicular structure in each bin. The peak of the Gaussian fit is taken to be the central optical depth.
 Then, in Figure \ref{vertical} we compare
the observed vertical structure to the expected vertical structure
for a filament and sheet as a function of distance from the protostar, converting
from the Gaussian $\sigma$ parameter to $H$ (for instance, we find that
$\sigma \sim 1.5 H$ for a filament).

For the case of L1157, the 10 K hydrostatic filament appears to be in reasonable
agreement with the observed extinction, while the sheet scale height
is about a factor of 3 than observed. If we do not
correct the 8$\mu$m extinction for foreground emission, 
the predicted filament scale heights were a
factor of 5 too large. 
As it is hard to imagine anything thinner than a pressure supported
filament, this result is further verification of the need 
for correcting the 8$\mu$m extinctions for foreground emission.

The envelope around Serp-MMS3 is also fit well by a filament over $\sim$0.1pc. 
We only fit the northeast part of the filament for this source because
the data of the southwest portion of the envelope is quite complicated.
In L673-SMM2, we attempted to fit both the north and south portions 
of the filament, avoiding the region near the protostars. 
The filament model does not fit as well as L1157 or Serp-MMS3; the observed
scale heights are always less than those predicted for a filament. 
We suspect that this discrepancy is due to 
the densest part of the filament being unresolved, underestimating the peak column density.

Neither sheet nor filament models yielded good fits to BHR71 which has a more complicated
envelope structure than L1157 and a larger optically thick region. 
The predicted scale height for a filament
tends to be about 2.5 times smaller than observed; thus a sheet seems to be more
consistent with the observed data. Alternatively, the dense structure may not be in 
hydrostatic equilibrium. 

\subsection{Mass Estimates}

With our optical depth maps at 
8.0 $\mu$m, we have the opportunity to estimate envelope masses 
independently of the temperature and chemical effects.  Our results do depend on
the assumed opacity, but recent extinction law studies using
 \textit{Spitzer} have yielded better constraints on the opacity at 8$\mu$m. However, we cannot trace
column density in regions where the envelope is optically-thick.

To derive a column density from the optical depths, we assume the dust plus gas opacity $\kappa_{8\mu m}=10.96$ cm$^2$ g$^{-1}$ \citep{butler2009} which 
corresponds to the \citet{weingart2001} $R_V$ = 5.5 Case B dust model which found to agree reasonably well
with the extinction laws derived at IRAC wavelengths \citep{roman2007,mcclure2009}. This opacity is calculated by 
convolving the IRAC filter response with the expected background spectrum shape, and opacity curve. 

To calculate the mass of a particular envelope, we simply sum all the pixels with $\tau$ greater than $\sigma\tau$ within  0.15, 0.1, and 0.05 pc
radii around the envelope and assume the relation
\begin{equation}
M_{env} =  d\Omega \times D^2 \times (1.496 \times 10^{13} \frac{cm}{AU})^2 \times \sum_{i}^{N} \frac{\tau_i}{\kappa}
\end{equation}
where $d\Omega$ is the pixel solid angle, (1.2$^{\prime\prime}$)$^2$, and $D$ is the distance in parsecs.

The masses determined using this method are probably lower limits at best because most 
envelopes become completely opaque in the densest areas. Moreover, the uncertainty in
optical depth increases  as optical depth increases. We are also probably not sensitive to
some mass on large scales due to signal-to-noise limitations and on small scales
emission is present which prevents measurement of optical depth.

The measured envelope masses
are given in Table 2 along with the measurements from ammonia and submillimeter studies.
The 0.05 pc radius is probably most comparable to the mass measurements from the other methods.
This is because most ammonia cores are generally about 0.1-0.15 pc in diameter and submillimeter
fluxes are generally measured in apertures with diameters of 80-120$^{\prime\prime}$ which correspond 0.1 - 0.15 pc
in diameter for an assumed distance of 300 pc. The masses we measure
are comparable with the other methods at smaller radii.

A specific trend that we see is that 
some protostars are surrounded by significant mass at large and small radii. All the protostars in the sample
are have nearly 1 M$_{\sun}$ of material within 0.05 pc; there is possibly even more mass within 0.05 pc since
we do no probe the regions where the protostar is emitting and the material could be optically thick.
Even the very low-luminosity protostars (e.g. IRAM 04191, L1521F, Perseus 5) are surrounded by many solar masses of material.
For reference we also list the bolometric luminosities for the observed sources in Table 2, there is no obvious
correlation between luminosity and envelope mass.

\section{Discussion}
The observations of dense, non-axisymmetric structures in Class 0 envelopes have significant implications
for our understanding of the star formation process. The dense structures that we observe either result from 
the initial conditions of their formation or they have been induced on an
otherwise axisymmetric envelope during the collapse process. We will discuss why these dense 
structures are not likely induced by outflows, the most obvious perturber, and then discuss how
non-axisymmetric infall could affect subsequent evolution of the system. We also discuss
how the envelope structures may give clues to the initial conditions of their formation and what relationship
the larger scale cloud structure around the protostar may have on its formation.

\subsection{Outflow-induced structure}

Outflows carve cavities in protostellar envelopes and necessarily affect the structure of the
protostellar environment at some level.  However, it is highly unlikely that 
outflows are responsible for much of the complex envelope morphology we observe.  
While a few objects such as L1527 exhibit relatively wide molecular outflows and
outflow cavities (e.g., \citet{jorgensen2007}), most of our outflow cavities 
(as seen most clearly in scattered light at $3.6 \mu$m, but also detectable in the  
8.0 $\mu$m images) are relatively narrow and well-collimated, in agreement with
molecular observations \citep[e.g.][]{arce2006,jorgensen2007}; this makes it difficult
to imagine that the outflow is strongly affecting most of the envelope.  Moreover,
if the observed dense, asymmetric envelope structures were the result of outflow sculpting, 
we would expect close spatial association of these structures with the outflow cavities; 
in contrast, the cavities in most objects are spatially distinct from the dense structures,
sometimes by quite large distances, and with no systematic alignment perpendicular to the cavities.

HH270 and IRAS 16253 are exceptions to this general rule, with strong concentrations
of extinction along cavity walls; however, both these systems are strongly asymmetric
perpendicular to the outflow cavities, which indicates that the density structures
are mostly the result of existing inhomogeneities in the ambient medium.
Similarly, it has been suggested that the outflow in L483 is strongly affecting
molecular material based on observations of N$_2$H$^+$ emission 
(\citep{jorgensen2004}); however, the strong density enhancement on one pole of the
outflow relative to the other is difficult to understand as purely a result of mass
loss, given the general bipolar symmetry of most systems.

To summarize, the outflows do not appear to have significantly affected the
current envelope morphologies in most cases.
The spatial separation of the outflow cavities and the dense, extincting structures make
it less likely that outflows can significantly limit mass accretion onto the central protostar.
The common inference that bipolar flows widen with age, dispersing the envelope
\citep[e.g.][]{arce2006,seale2008} may result more directly from collapse of dense
structures rather than a changing of the outflow angular distribution. 

\subsection{Dense Non-axisymmetric Structure}

The shapes of dense cores have been studied on large scales ($>$0.1 pc) 
using optical extinction \citep{ryden1996} and molecular line tracers of dense gas
\citep{benson1989,myers1991,caselli2002}. However, optical extinction
only traces the surface of dense clouds and most envelopes appear
round in the dense molecular tracers because they are generally only
spanned by 2.5 beams or less. The low-resolution molecular line studies are
slightly more advantageous compared to the optical since they only detect
dense material and associated IRAS sources are often located off-center
from the line emission peaks indicating non-axisymmetry.

The advantage of 8$\mu$m extinction in studying envelopes compared to other methods
is that it provides high resolution, undiminished sensitivity to dense extended 
structures, and its a tracer that depends only on density. This enables us to 
trace non-axisymmetric structure from large scales down to 1000 AU scales.
Figure \ref{CB230} exemplifies the details revealed by 8$\mu$ extinction
maps; the 8$\mu$m extinction contours of CB230 are overlaid
on the optical Digitized Sky Survey image showing the strong asymmetry of dense
material while the optical image shows no hint of what is going on at small scales.
The necessity of 8$\mu$m extinction to see small scale non-axisymmetric structure
holds true for all our sources. The magnitude non-axisymmetry varies significantly 
between sources; but the important point is that all sources exhibit non-axisymmetry
and that sources with ``regular'' morphology are the exception rather than the rule.

CB230 and the other one-sided envelopes shown in Figure \ref{oneside} are particularly
intriguing. While the outflows may have done some sculpting, the dynamics of the
star formation process itself has caused the strong asymmetries perpendicular to the outflow.
The curvature of dense structures in BHR71 and L723 as well as the outflow being oriented
non-orthogonal to many dense filamentary structures (e.g., L673, HH108, SerpMMS3, L1448 IRS2)
may indicate that the angular momentum of a collapsing system may not have a strongly preferred
direction set by the large scale cloud structure.

It is important to point out that these non-axisymmetric structures exist down to small scales
quite near the protostar and only at about $\sim$1000AU they become obscured by emission from the
protostar 8$\mu$m. As shown in Table 2, all the envelopes have a nearly 1 M$_{\sun}$ or more
 mass within 0.05 pc. Material we see at $\sim$1000AU potentially has an infall timescale
of $\sim$5$\times$10$^4$ years and at 10000 AU the infall timescale is about $\sim$5$\times$10$^5$ years
 assuming a 0.96 M$_{\sun}$ initial core at the start of collapse
(\S3 \citep{shu1977}). This is consistent with Class 0 protostars
having accumulated less than about half their final mass at the present
epoch \citep{myers1998, dunham2010}.

\subsection{Implications of Non-Axisymmetric Collapse}

The envelope asymmetries we see may well result from the initial cloud structure.
\citet{stutz2009} recently surveyed pre-stellar/star-less cores using 8 and 24$\mu$m
extinction; their results, and those of \citet{bacmann2000}, showed that even
pre-collapse cloud cores already exhibit some non-axisymmetry. Given the initial asymmetries
the densest, small-scale regions are likely to become even more anisotropic during
 gravitational collapse \citep{lin1965}. We note that gravity is not the only force at work
in these clouds; turbulence and magnetic fields may also play roles in forming
the envelope morphologies (\S6.4). 

The smallest scales we observe, $\sim$1000 AU, is where angular momentum will begin
to be important as the material falls further in onto the disk.
The envelope asymmetries down to small scales imply that infall
to the disk will be uneven; therefore,  non-axisymmetric infall
may play a significant role in disk evolution and the formation of binary systems. 
Several theoretical investigations \citep[e.g.][]{burkert1993, boss1995} 
showed that collapse of a cloud with just a small azimuthal perturbation can
form binary or multiple systems; thus, {\em large} non-axisymmetric perturbations
should make fragmentation even easier. Fragmentation can even begin before
global collapse in a filamentary structure \citep[][and references therein]{bonnell1993}.
Numerical simulations of disks with infalling envelopes 
\citep[e.g.][]{kratter2009,walch2009} informed by the results of this study could reveal 
a more complete understanding of how non-axisymmetric infall affects the disk and infall process.

Several sources in the sample are known wide binaries (BHR71, CB230) \citep{bourke2001, launhardt2001}
 and close binaries (L1527,
L723, IRAS 03282+3035) \citep{loinard2002, girart1997, chen2007}. The \textit{Spitzer} observations
 indicate that SerpMMS3 may be a binary and that L673-SMM2 is likely a multiple.
Other sources in our sample may be close binaries but this property can only be revealed
by sub-arcsecond imaging. \citet{looney2000} showed that many
protostars are indeed binary when viewed at high enough resolution at millimeter wavelengths.
A recent study of Class 0 protostars at high resolution by \citet{maury2010} noted that
their results taken with \citet{looney2000} show a lack of close binary systems
with 150-550AU separations. This may signify that non-axisymmetric infall throughout the Class 0
phase is important for binary formation later when centrifugal radii extend out to 150-550AU.

\subsection{Turbulent formation}

The dense, non-axisymmetric structures that we observe around Class 0 protostars
are not obviously consistent with quasi-static, slow evolution,
which might be expected to produce simpler structures as irregularities have
time to become damped. With rotation, one might get a flattened system
during collapse, as is well known \citep{tsc1984}, but one needs non-axisymmetric initial
structure to get strong non-axisymmetric structure later on. This raises the question
of the role of magnetic fields in controlling cloud dynamics. In some models
\citep[e.g.][and references therein]{fiedler1992,galli1993, tassis2007,kunz2009}, protostellar
cores would probably live long enough to adjust to more regular configurations; in addition,
collapse would be preferentially along the magnetic field, which would 
also provide the preferential direction of the rotation axis and therefore for the (presumably magnetocentrifugally
accelerated) jets \citep{basu1994, shu1994}. The
complex structure and frequent misalignment between collapsed structures
and outflows pose challenges for such a picture.

In contrast, more recent numerical simulations
\citep[e.g.][]{padoan2001,klessen2000, ballesteros1999} suggest that cores are the result
of turbulent fluctuations which naturally produce more complex structure
with less control by magnetic fields amplified by subsequent gravitational contraction and collapse 
\citep[e.g.][see review by Ballesteros-Paredes et al. 2007 (PPV)]{elmegreen2000,klessen2000,
klessenburkert2000,heitsch2001,padoan2002,hartmann2002,bate2003,maclow2004,heitschhartmann2008,
heitsch2008a,heitsch2008b}. Thus, the structure of protostellar envelopes
thus provides an indication of which of the two contrasting
pictures of core formation, with differing assumed timescales of formation
and differing importance of magnetic fields, is more nearly correct.

The timescale argument against ambipolar diffusion can be mitigated since turbulence is known to 
accelerate ambipolar diffusion \citet{fatuzzo2002,basu2009}. 
The filamentary envelopes may also enhance
ambipolar diffusion as necessary. For example, a cylinder with an aspect ratio of 4:1
(similar to L1157) will have a factor of $\sim$ 10 less volume 
than a sphere with a diameter equal to the cylinder length.
If both have the same infall rate,
the filament has a factor of $\sim$ 10 higher density than the sphere. Thus, using
\begin{equation}
\label{adtimescale}
\tau_{AD} \sim 5 \times 10^{6}\left(\frac{10^4 cm^{-3}}{n(H_2)}\right)^{\frac{1}{2}} yr
\end{equation}
given in \citet{spitzer1968}, the ambipolar diffusion timescale, $\tau_{AD}$, is reduced by an order of magnitude!
With a shorter ambipolar diffusion timescale, the magnetic support 
of the initial disk could be lower than the levels suggested in \citet{galli2006}, allowing
for Keplerian rotation of the resulting circumstellar disk.  

The timescale issue aside, it is still difficult to get non-axisymmetric
structures from magnetic collapse. Simulations by \citet{basu2004, ciolek2006}
indicated that magnetically sub-critical or
critical cores will tend to be round or axisymmetric while
the super-critical cores would show higher degrees of non-axisymmetry.
Our results are more consistent with fast, super-critical collapse.
Further high resolution simulations, such as those by \citet{offner2009}, 
would help make a better connection between theory and observation. 

Alternatively, if a spherical core forms within a turbulent medium, as 
in \citet{walch2009}, the turbulence within the core itself could give rise
to a non-axisymmetric structure.  However, two difficulties of this scenario
are immediately obvious; first, producing a spherical core in a turbulent
environment seems difficult; and second, dense, star forming cores are found to be very quiescent
compared to their external medium \citep{goodman1998,pineda2010}. Rather than
the core itself being turbulent, anisotropies in the turbulent pressure
surrounding a dense core could also give rise to non-axisymmetric structure
from an initially symmetric core. One could also envision a scenario where
an envelope is impacted by colliding flows causing non-axisymmetries.

\subsection{Relationship with Larger Structures}

While some systems in our sample seem to be in relative isolation, 
most are part of large-scale filamentary structure. With the dataset
presented, we can examine the spatial relationship of the protostars within
their natal material to see if there are trends which influenced
by the non-axisymmetries. Here we examine several of the protostars where we
can clearly discern the morphology of larger scale material.

The L1165 dark cloud is comprised of a long filament running from southeast to northwest for $\sim$8$^{\prime}$ (0.6pc)
that turns northeast forming a roughly 90$^{\circ}$ angle and extends $\sim$10$^{\prime}$ (0.75pc).
An image of the L1165 region at 8$\mu$m and a near-IR extinction map are shown in Figure
\ref{L1165}.  Both the protostar (IRS1) and another very bright source about 1.5$^{\prime}$ north (IRS2) have formed
near the `elbow' of the filament. IRS2 is likely a young star because the spectral index from 6-13$\mu$m is $\sim$-1
indicating that it is a Class II object. IRS2 is detected at 24$\mu$m (fainter than IRS1) but not at 70$\mu$m.
It is intriguing that these two young stars have formed at the  `kink' in the filament while there are no apparent protostars
 elsewhere in the filament. Two other examples of multiple stars forming at filament kinks
are L673-SMM2 and Serp-MMS3.

The protostars HH108 IRS1 and HH108 IRS2, (IRAS and MMS respectively in \citet{chini2001}) have
also formed with a filament. HH108 IRS1 is more luminous and IRS2 only appears in emission longward of 24$\mu$m.
IRS2 appears as an opaque spot in the 24$\mu$m extinction image in Figure 1. The envelope of IRS1 appears to be
a collapsed portion of the larger filament and is located at a bend.
On the other hand, the envelope of IRS2 appears round, but slightly extended along the filament.

The protostars CB230 and IRAS 03282+3035 are both located at the edges of dark clouds. We 
can clearly see the edge of the envelope around IRAS 03282+3035 in diffuse scattered light
at 3.6$\mu$m corresponding to the edge of 8$\mu$m extinction (Figure 3). 
CB230 is an isolated Bok globule, shown in Figure \ref{CB230}. The dark cloud in the optical extends to the west
from the protostar $\sim$11$^{\prime}$ (0.95 pc) with highest densities near the protostar at the extreme eastern
edge of the cloud. The optical depth map shows the extreme asymmetric distribution of material.
In addition, there is an optical star associated with a reflection nebula $\sim$7.25$^{\prime}$ west of the protostar,
but it is not detected by IRAS and not observed by\textit{Spitzer}.

The L1448 dark cloud contains several embedded protostars visible at 8$\mu$m \citep{tobin2007}.
The most isolated protostar is L1448 IRS2, the rest are surrounded by bright emission from
outflow knots. IRS1 and IRS2 are located toward the western edge of the cloud
IRS3A/B are located on the northeastern corner of the cloud and L1448-mm is located in the southeast corner.
There is a filament of material running between IRS2 and IRS3 seen in diffuse scattered light
(Figure 1 in \citet{tobin2007}), 8$\mu$m extinction,and ammonia emission \citep{anglada1989}.
Also, the filament abruptly cuts off $\sim$40$^{\prime\prime}$ north of IRS2.

Protostars highlighted seem to have a tendency to for
to form at the edges of clouds or where there are turns or 'kinks' in a
filamentary structure. This behavior is qualitatively what would be expected if gravitational
focusing is important. Put simply, gravitational
focusing causes the edges of clouds and where there are discontinuities (bends and
kinks) to form stars first by creating gravity focal points. In a filament undergoing global
collapse, the ends of that filament will be moving inward the fastest and will encounter slower moving material. This 
`gravitational traffic jam' creates the gravity focal point. In addition, the filament will also be collapsing
 in the vertical direction which causes material to flow toward the focal point in from orthogonal directions
rather than just the transverse direction. A scenario such as this would cause the ends of the filament to form stars
rather than at the center of the filament. Quantitatively, this scenario appears when you modify the
spherical (3D) free-fall timescale for a (1D) filamentary geometry.
Gravitational focusing is seen in theoretical work by \citep{burkert2004} which simulated a complex object
with many thermal Jeans masses. The gravitational
focusing causes non-linear collapse near cloud boundaries and
other discontinuities (see also Bonnell \& Bastien 1993).  The results are consistent
with a picture in which turbulent fragmentation provides ``seeds'' which then are
amplified by gravity \citep{heitsch2008b, heitschhartmann2008}.

We note, however, that BHR71 and L483 seem 
to have formed at the centers of centrally condensed Bok globules,
so we cannot claim complete universality for this mechanism. However, the filamentary regions
 highlighted in the above paragraphs may be representative of star forming environments because
the theoretical work shows that turbulent star formation generally gives rise to filaments 
\citep[e.g.][]{heitsch2008b, heitschhartmann2008}. Also, the preference of forming stars in clumps at the ends of 
filaments also appears to apply to young star clusters and loose star-forming associations 
(e.g. Orion Nebula Cluster, NGC2264, Chameleon, Taurus)
\citep{furesz2006,furesz2008,tobin2009,luhman2009,hartmann2002}. Thus, gravitational focusing may be at work 
from the formation of clusters to the formation of individual protostars. Further theoretical work building on
that of \citet{burkert2004} including effects of turbulence and/or magentic fields would enable a better understanding
of this idea and determine how much gravity must dominate over other forces present in the cloud.

\section{Conclusions}

In this paper, we have shown the complex structure of envelopes surrounding Class 0 protostars as viewed
in extinction at 8$\mu$m using \textit{Spitzer} IRAC images. The non-axisymmetries revealed
by the IRAC extinction maps were not obvious in submillimeter maps by SCUBA. The 8$\mu$m images were found
to be significantly contaminated by foreground emission and we corrected for this using near-IR
extinction measurements toward background stars. This method demonstrated that the densest parts
of the envelopes observe are completely opaque at the observed signal-to-noise levels.
We have characterized the non-axisymmetry of the envelopes in terms of
projected moments of inertia ratios. Most envelopes are more extended perpendicular
to the outflow, but there are exceptions. Our measurements also yield estimates of the mass
surrounding the protostars at small and large scales.

Most envelopes show highly non-axisymmetric structure from $\sim$1000AU to 0.1 pc scales.
These asymmetric structures are \textit{not} caused by outflow-envelope interactions
as the outflows are still highly collimated and spatially located away 
from asymmetric structures and envelopes tend to be more extended perpendicular
to the outflow. We suggest that the widening of outflows with age may result more directly from collapse of dense
structures rather than a changing of the outflow angular distribution.
In the entire sample, we find significant mass present in the envelopes 
on scales less than 0.05 pc. This supports the idea
that Class 0 protostars are in the main phase of mass accretion and the asymmetries of the
material down to small scales indicates that material will likely fall onto the disk unevenly
possibly enhancing gravitational fragmentation. This could
help explain the formation of close binary or multiple systems.

The highly non-axisymmetric envelopes may result directly from
the collapse of mildly asymmetric cores found by \citet{bacmann2000,stutz2009}
because fast collapse will enhance anisotropies \citep{lin1965}; turbulence
and colliding flows could also play a role in creating asymmetries.
The observed structure points to non-axisymmetric and probably
non-equilibrium initial conditions. If the magnetic field plays a major role
in the collapse process of these envelopes, it is clearly not working to make the infall
process more symmetric. Comparison to simulations indicates that super-critical
collapse is more consistent with our observations.

Finally, there seems to be a preference of where stars form within larger scale structures.
Several systems form protostars at the ends of filaments and at bends or kinks 
in the more-extended molecular gas,
which suggests that the initial shape of a cloud has much to do with where stars form. This
is reminiscent of the preference for stars to form in clusters double clusters as we see
in local star forming regions such as Orion, NGC2264, and Chameleon.

We thank the anonymous referee for comments enabling us to improve the clarity and impact
of this paper significantly. We wish to thank F. Heitsch, Y. Shirley, and A. Stutz
for useful discussions and insight. We thank W. Fischer and J. Hernandez for
conducting the PANIC observations while J. Tobin was on his honeymoon.
We are grateful to the staff of MDM Observatory  (R. Barr, J. Negrete, and P. Hartmann) 
for their support during TIFKAM observations. We thank N. van der Bliek for her support
during the ISPI observing run. The observing time on the Blanco 4m of CTIO
was made possible by a partnership between the University of Illinois and NOAO.
TIFKAM was funded by The Ohio State University, 
the MDM consortium, MIT, and NSF grant AST-9605012. The HAWAII1 array used 
in TIFKAM was purchased with an NSF Grant to Dartmouth University.
 This publication makes use of data products from the Two Micron All Sky Survey, which
is a joint project of the University of Massachusetts and the Infrared
Processing and Analysis Center/California Institute of Technology,
funded by the National Aeronautics and Space Administration and the
National Science Foundation. J. J. T. and L. H. acknowledge funding 
from the \textit{Spitzer} archival research program 50668;
NASA grant 1342979. L. W. L. and H. C. acknowledge support from
the National Science Foundation under Grant No. AST-07-09206 and \textit{Spitzer} GO program 30516.

{\it Facilities:} \facility{Spitzer (IRAC)}, \facility{Hiltner (TIFKAM)}, \facility{Magellan:Baade (PANIC)}, \facility{Blanco (ISPI)}

\section{Appendix}

As discussed in \S4, it was necessary to correct our data for foreground emission.
Some analyses have used submillimeter emission maps to correct for foreground
emission \citep{johnstone2003} and \citet{ragan2009}.
These analyses assume that the
absorbing material has become optically thick 
at the points where submillimeter emission is increasing
but the IRAC 8$\mu$m intensity reaches its minimum and
the 8$\mu$m intensity at that position is taken to be the foreground emission.
However, we cannot apply this method to the envelopes 
in our study because the protostar warms its envelope making the submillimeter 
emission dependent on both temperature and density. 

Instead, we compared the near-IR extinction of background stars (viewed through the envelopes) 
to the 8$\mu$m extinction enabling the determination of the foreground emission. 
The main difficulty in applying this method is that the envelope must be viewed against a 
rich stellar background to enable an accurate determination of $I_{fg}$.
In addition, the near-IR image must be deep enough to measure accurate photometry
in at least H and Ks bands; data from the 2MASS survey are too shallow. 
to the best of our knowledge, this is the first time this method has been employed to constrain
the amount of diffuse foreground emission in the construction of extinction maps from extended
emission.

We start by assuming that the foreground emission
can be taken as a constant offset to the true background by
\begin{equation}
I_{obs,bg}^{\prime}=  I_{obs,bg} + I_{fg}
\end{equation}
where $I_{obs,bg}^{\prime}$ is the background intensity or observed intensity in an extincted region
which has some constant $I_{fg}$ present due to the possible effects listed in the previous paragraph.
The presence of foreground emission changes the the optical depth relationship of Equation (\ref{taucalc})
to be
\begin{equation}
\label{tauuncorr}
\frac{I_{obs}^{\prime}}{I_{bg}^{\prime}}=\frac{I_{obs} + I_{fg}}{I_{bg} + I_{fg}}=e^{-\tau_8^{\prime}}
\end{equation}
where $\tau_8^{\prime}$ 
is the measured optical depth from the IRAC images that are not corrected
for foreground emission. As $I_{fg}$ increases, the ratio of 
$I_{obs}^{\prime}/I_{bg}^{\prime}$
increases, causing the measured optical depth to decrease.

The NICE method \citep{lada1994} enabled us to measure the extinction toward stars by assuming the background
stars can be described by a single, average color. The extinction is determined from
\begin{equation}
\label{nicefull}
A_H - A_{Ks} = [(H - Ks)_{obs} - \langle H - Ks\rangle_{off}] = A_{Ks}\left(\frac{A_H}{A_{Ks}} -1\right)
\end{equation}
where $(H - Ks)_{obs}$ is the color of an individual star, while $\langle H - Ks\rangle_{off}$
 is the mean color of the background stellar population. 
$A_H/A_{Ks}$ is known from near-IR extinction law measurements to be 
$\sim$1.56 \citep[e.g.][]{indeb2005,rieke1985}. This value can vary between 1.5-1.6 for the different possible power-law dependencies of
the near-IR extinction law which assumes $A_{\lambda} \propto \lambda^{- \beta}$ and $\beta$ is known
from observation and dust models to be between 1.6 and 1.8 \citep[e.g.][]{weingart2001}. The value we assume from \citet{rieke1985} 
has $\beta$ = 1.71. The result is 
\begin{equation}
\label{nicecalc}
A_{Ks} = 1.77\times[(H - Ks)_{obs} - \langle H - Ks\rangle_{off}].
\end{equation}

We determined $\langle H - Ks\rangle_{off}$ by using the 2MASS 
catalog to calculate the $(H - Ks)$ color of stars near the protostar but estimated to
be relatively free of extinction, 
as judged from visual inspection of the
the optical DSS images. Then we create a histogram of $(H - Ks)$ colors
and fit a Gaussian to the distribution.
The mean is then taken to be the value for $\langle H - Ks\rangle_{off}$; this
value is generally $\sim$0.2.
Then for each star we have a measure of extinction at Ks-band, $A_{Ks}$. The value of $A_{Ks}$ 
is uncertain for an individual star; therefore we compare 
$A_{Ks}$ to $A_{8\mu m}$ at many points throughout the envelope.
The appropriate extinction laws \citep{flaherty2007,roman2007,mcclure2009}
indicate that $A_{8\mu m} = 0.5 \times A_{Ks}$ for most star forming regions. We
can then extrapolate the optical depth at 8$\mu$m from near-IR extinction ($\tau_{8,Ks}$) to be
\begin{equation}
\label{taufromks}
\tau_{8,Ks} = (1.068)(0.5) \times A_{Ks}
\end{equation}
where $A_{Ks}$ is determined from the near-IR extinction measurement.
Figure \ref{taucompnocorr} shows the uncorrected relationship between $A_{Ks}$ to $A_{8\mu m}$ for BHR71 and L483.
The deviation of the predicted relationship from the observations clearly illustrates the necessity of applying this
method to our sample to determine the level of foreground contamination.

Figure \ref{taucompnocorr} clearly indicates that our initial optical depth measurements 
from Equation (\ref{tauuncorr}) needed to be corrected for extra emission.
Our near-IR extinction analysis yields the true optical depth from 
\begin{equation}
\label{taucorr}
\frac{I_{obs}}{I_{bg}} = e^{-\tau_{8,Ks}}
\end{equation}
which is equivalent to
\begin{equation}
\label{taucorrprime}
\frac{I_{obs}^{\prime} - I_{fg}}{I_{bg}^{\prime} - I_{fg}} = e^{-\tau_{8,Ks}}.
\end{equation}
Solving for $I_{fg}$ and some algebraic manipulation give
\begin{equation}
\label{fgcalc}
I_{fg}=\frac{I_{bg}^{\prime}(e^{-\tau_8^{\prime}}-e^{-\tau_{8,Ks}})}{1-e^{-\tau_{8,Ks}}}.
\end{equation}
This relationship assumes that $I_{fg}$ is nearly constant across the envelope, which 
is a reasonable assumption for the possible sources of foreground given the relatively small angular size of the envelopes. 

We have already described how $I_{bg}^{\prime}$ is measured in the previous section. However, $\tau_8^{\prime}$ from Equations \ref{tauuncorr}
and \ref{fgcalc} requires some special consideration. 

Since the near-IR extinction toward these points is determined using stars, there may
be point sources detected at the same position in the 8$\mu$m image which will have a negative value in the
optical depth map. This is particularly clear in Figure \ref{nirimages} where some background stars are surrounded
by a ``hole'' in the optical depth map. To ensure that our measurements of $\tau_8^{\prime}$ from the 8$\mu$m extinction map are mostly
unaffected by stars, we measure the average optical depth in an annulus from 4 to 7 pixels around the star at each position. 
This does introduce some error if a star
is very bright at 8$\mu$m, seen in Figure \ref{taucompcorr} as points with high $A_{Ks}$ but low $A_{8\mu m}$,
but for most points in the envelope this method works well. Additional points with high $A_{Ks}$ and low $A_{8\mu m}$
are likely due to the presence of diffuse emission at 8$\mu$m from scattered light in the outflow cavity, outflow knots, and/or
a star(s) falling within the measurement annulus. As a test, we compared the average $A_{Ks}$ in a 15$^{\prime\prime}$
box to the median 8$\mu$m extinction in the same box and the points with high $A_{Ks}$ and low $A_{8\mu m}$ were not present.

We determine $I_{fg}$ by calculating Equation (\ref{fgcalc}) at the position of each near-IR extinction measurement  where $A_{Ks} > 1$ and
$A_{8\mu m}$ is greater than the 1 sigma noise in the uncorrected optical depth map. Then we take the median value
of $I_{fg}$ and subtract it from the 8$\mu$m image and recalculate the optical depth map. Then we run the
comparison again on the corrected image and $I_{fg}$ should be close to zero; graphically  we check to see if the datapoints agree
with the predicted relationship, usually $I_{fg}$ will only need to be adjusted slightly from the initial value. We note that the 
regions of highest $\tau_8^{\prime}$/$A_{Ks}$ yield the best value of $I_{fg}$ because the percentage error
for these points is the least; there can be significant scatter at low $\tau_8^{\prime}$/$A_{Ks}$.
As shown in Figure \ref{taucompcorr}, the correction to the optical depth 
measurements results in reasonable estimates of the foreground contribution. As highlighted
in \S 4 this method lead us to the simplifying conclusion that in all images the darkest
part of the envelope is completely opaque within uncertainty limits of the images. This 
finding enables us to also correct our data which lack near-IR measurements or are
not observed against a rich stellar field.

\begin{small}
\bibliographystyle{apj}
\bibliography{ms.v3}

\begin{figure}
\begin{center}
\includegraphics[scale=0.7]{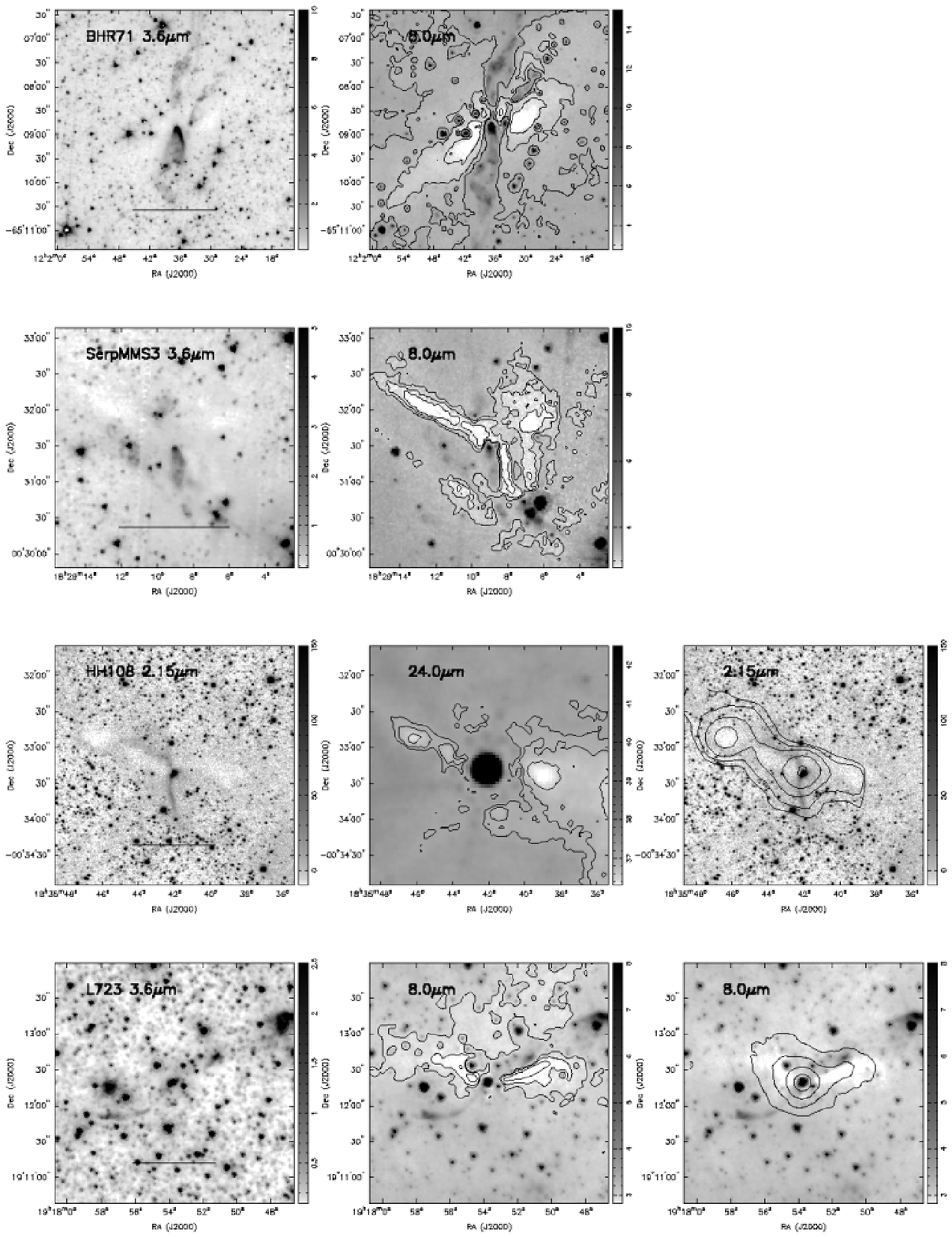}
\end{center}
\caption{IRAC images of our sample of envelopes with flattened morphology. Left: IRAC 3.6 $\mu$m or Ks-band images
which highlight the scattered light cavities in these objects. Middle: IRAC 8.0 $\mu$m images
with the 8.0 $\mu$m optical depth contours overlaid. Right: IRAC 8.0 $\mu$m images with
SCUBA 850 $\mu$m contours. The line drawn in the left
panel corresponds to 0.1 pc at the adopted distance. For HH108, we show the Ks-band image (left), 24$\mu$m image
with 24$\mu$m optical depth contours (middle) highlighting structure, and SCUBA 850$\mu$m data
overlaid on the Ks-band image (right). The 8.0 $\mu$m optical depth contours
correspond to $\tau_{8\mu m}$ values for BHR71: 0.6, 1.28, 2.75; L723: 0.2, 0.3, 0.45; L673: 0.3, 0.6, 1.2; SerpMMS3: 0.75 1.06 1.5;
L1157: 0.5, 1.09, 2.4.
}
\label{flat}
\end{figure}

\clearpage
\centerline{\includegraphics[scale=0.7,angle=-90]{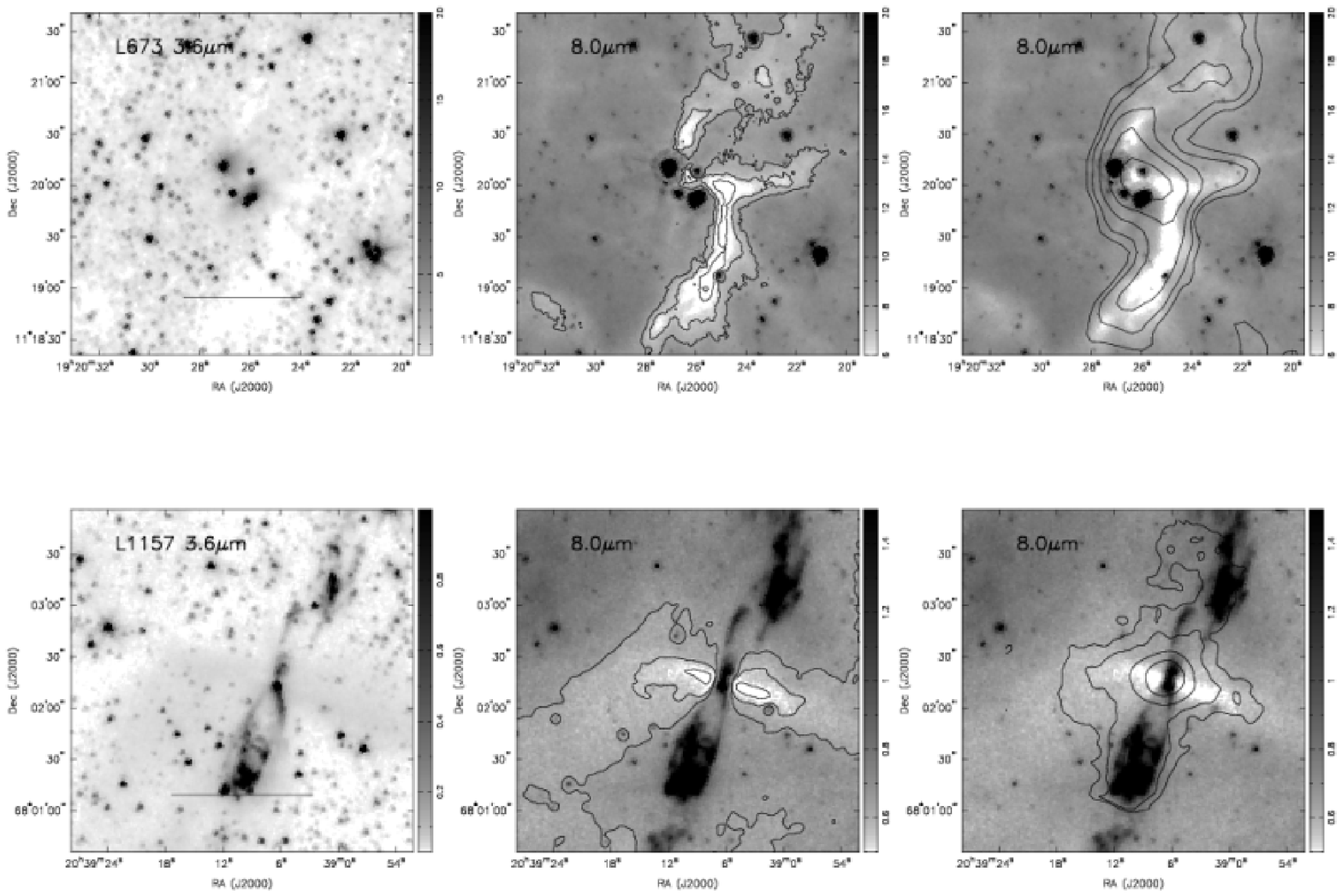}}
\centerline{Fig. 1 ---}

\begin{figure}
\begin{center}
\includegraphics[scale=0.8]{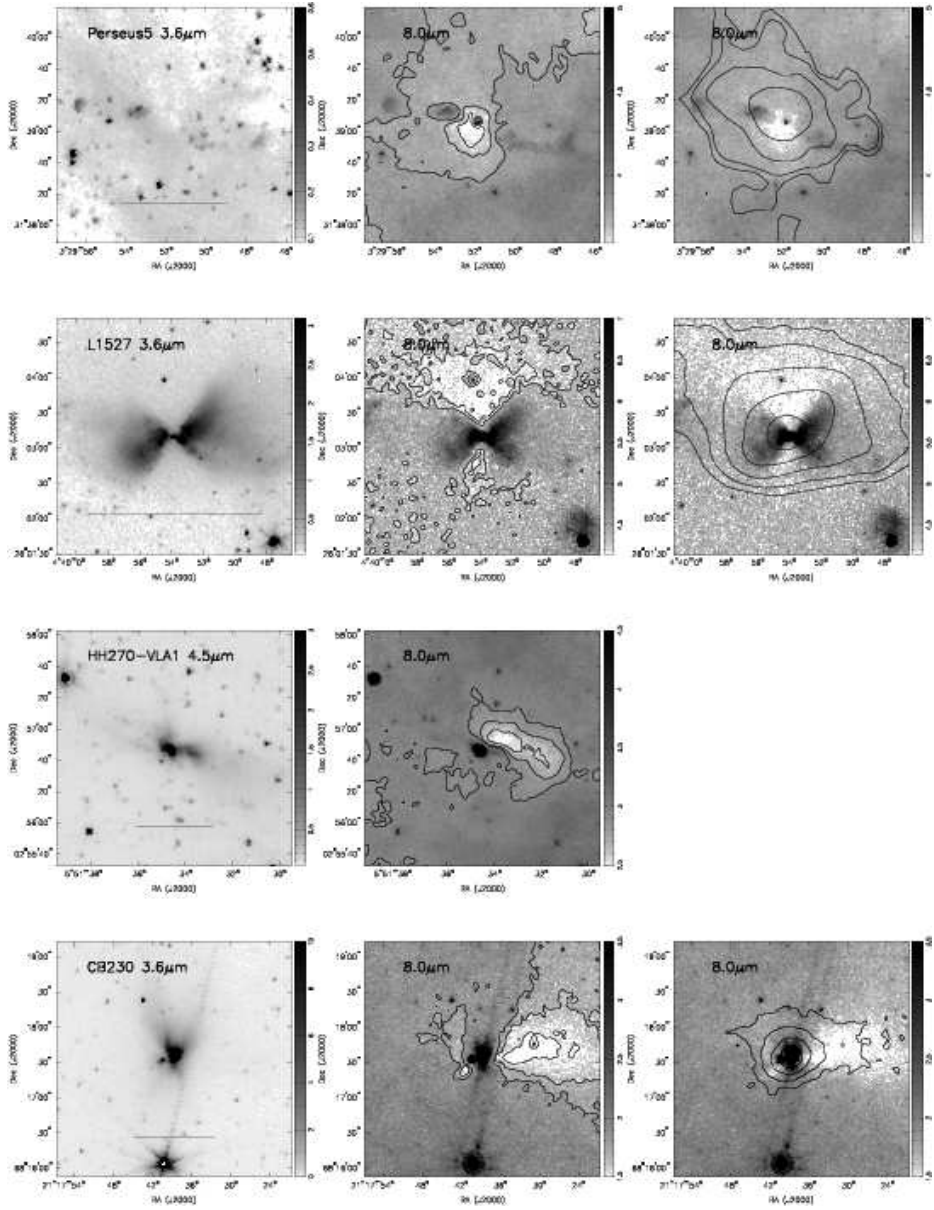}
\end{center}
\caption{Same as Figure 1 except one-sided envelopes are shown. The 8.0 $\mu$m optical depth contours
correspond to the following values of $\tau_{8\mu m}$ for Perseus 5: 0.75, 1.22, 2.0; L1527: 0.1 0.375, 1.4;
HH270 VLA1: 0.34, 0.6, 2.4; CB230: 0.3, 0.67, 1.5.}
\label{oneside}
\end{figure}

\begin{figure}
\begin{center}
\includegraphics[scale=0.8]{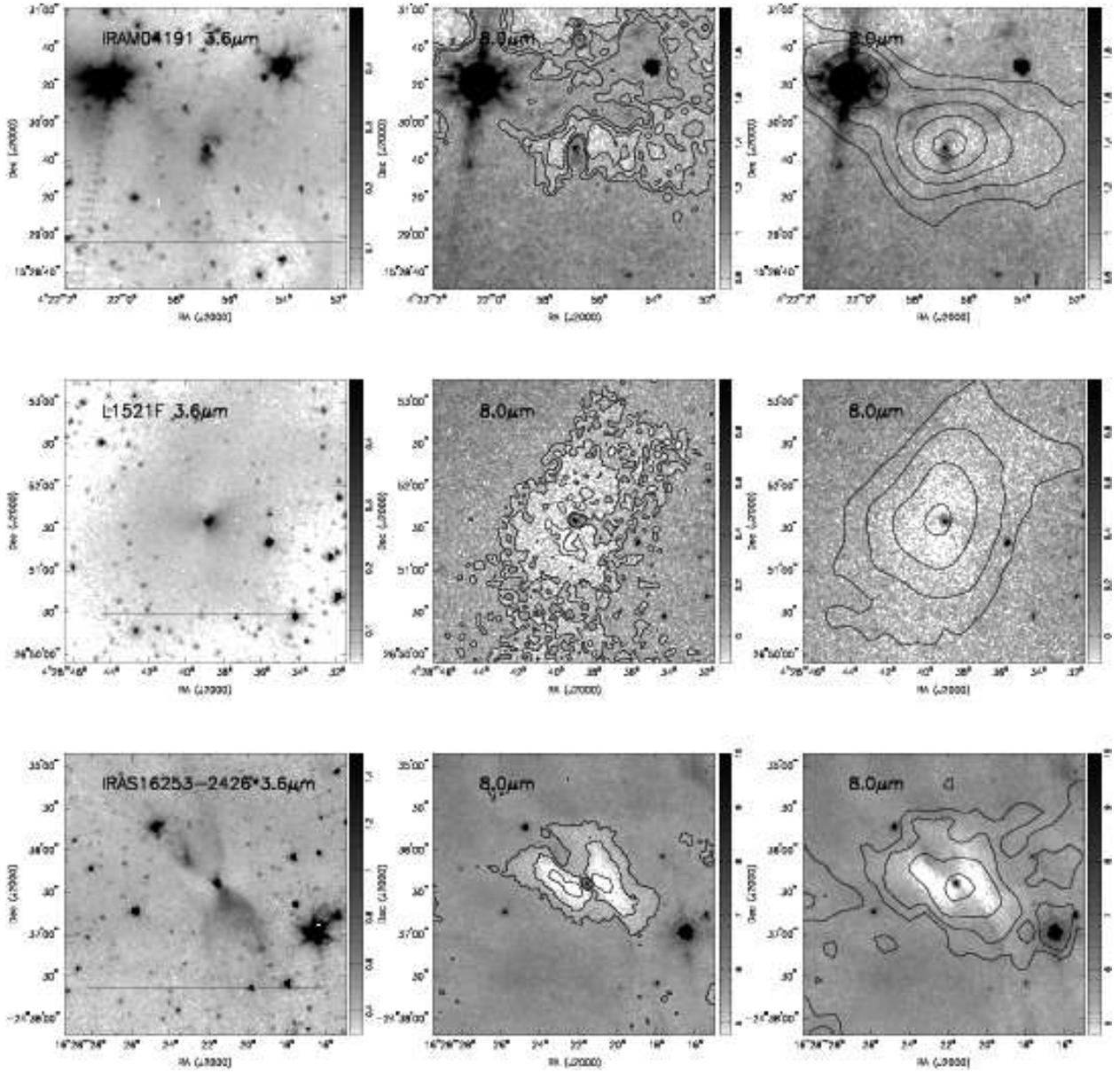}
\end{center}
\caption{Same as Figure 1 except spheroidal envelopes are shown. The 8.0 $\mu$m optical depth contours
correspond to the following values of $\tau_{8\mu m}$ for IRAM 04191: 0.4, 0.57, 0.8; L1521F: 0.6, 0.85, 1.2;
IRAS 16253-2429: 0.3, 0.53, 0.95.}
\label{round}
\end{figure}

\begin{figure}
\begin{center}
\includegraphics[scale=0.6]{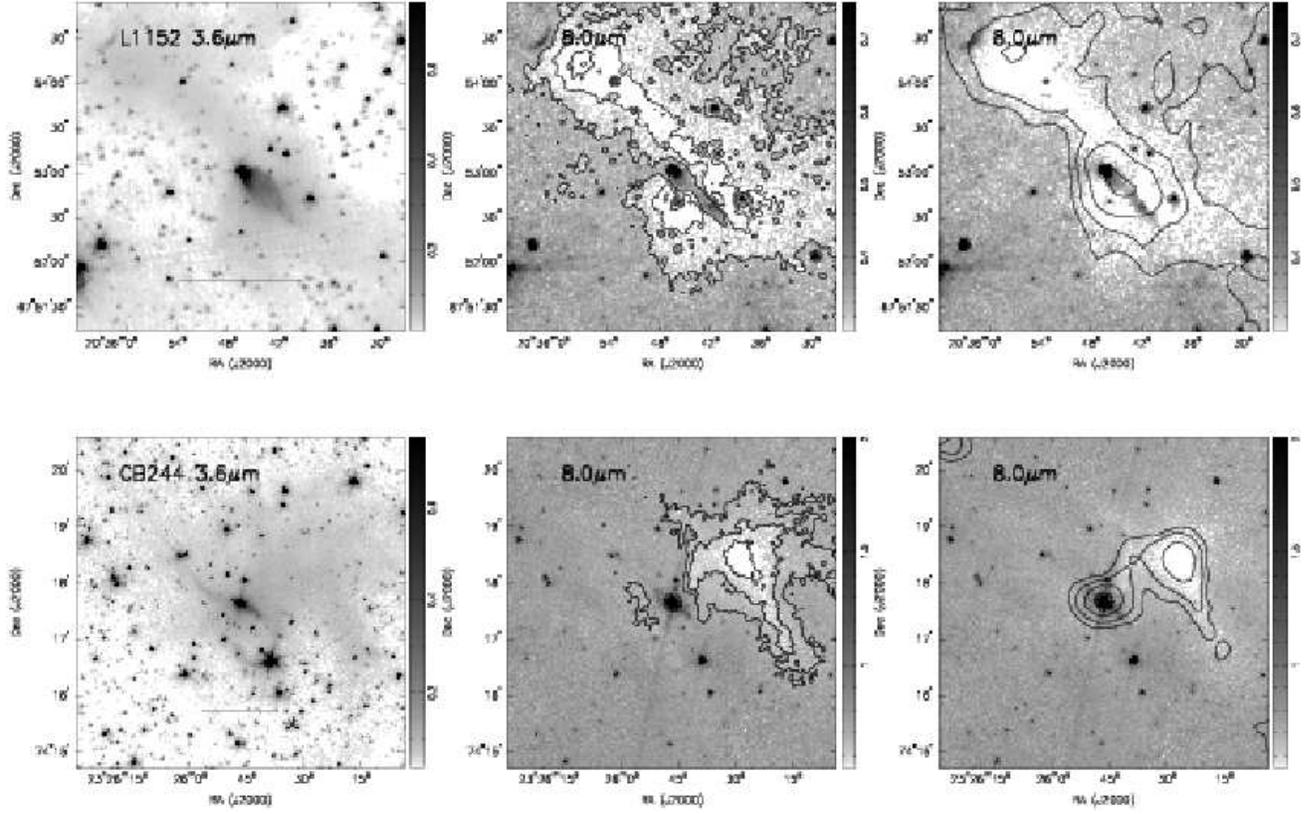}
\end{center}
\caption{Same as Figure 1 except binary envelopes are shown. The 8.0 $\mu$m optical depth contours
correspond to the following values of $\tau_{8\mu m}$ for CB244: 0.3, 0.72, 1.75; L1152: 0.45, 0.91, 1.85.}
\label{binary}
\end{figure}

\begin{figure}
\begin{center}
\includegraphics[scale=0.8]{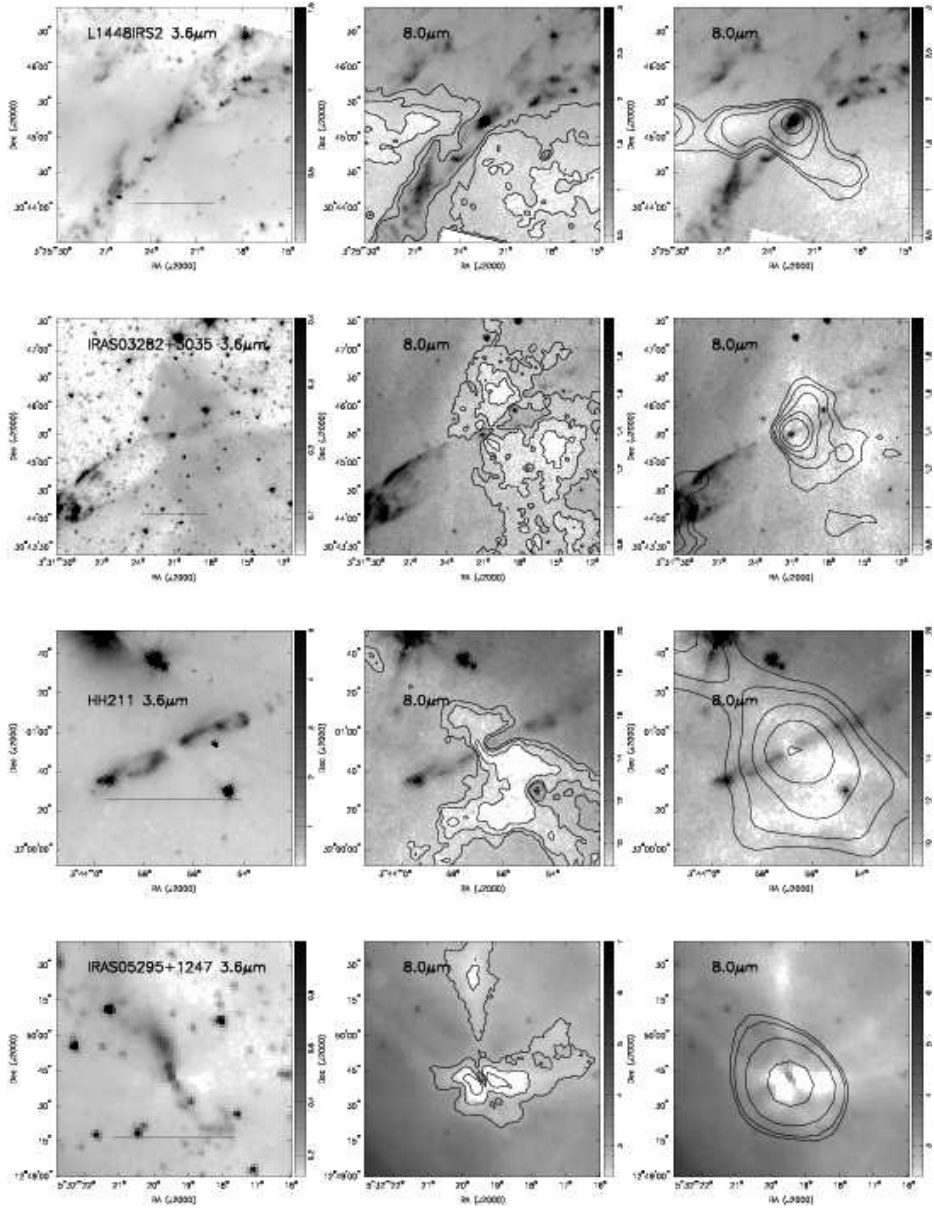}
\end{center}
\caption{Same as Figure 1 except irregular envelopes are shown. The 8.0 $\mu$m optical depth contours
correspond to the following values of $\tau_{8\mu m}$ for L1448 IRS2: 0.6, 1.02, 1.75; 
IRAS 03282+3035: 0.9, 1.34,2.0; HH211 0.3, 0.57, 1.1; IRAS 05295+1247 0.275, 0.74, 2.0; IRS09449-5052: 0.25, 0.56, 1.25; L483: 0.5, 0.82, 1.35; 
L1165: 0.3, 0.54, 1.0.}
\label{irregular}
\end{figure}

\clearpage
\centerline{\includegraphics[scale=0.8]{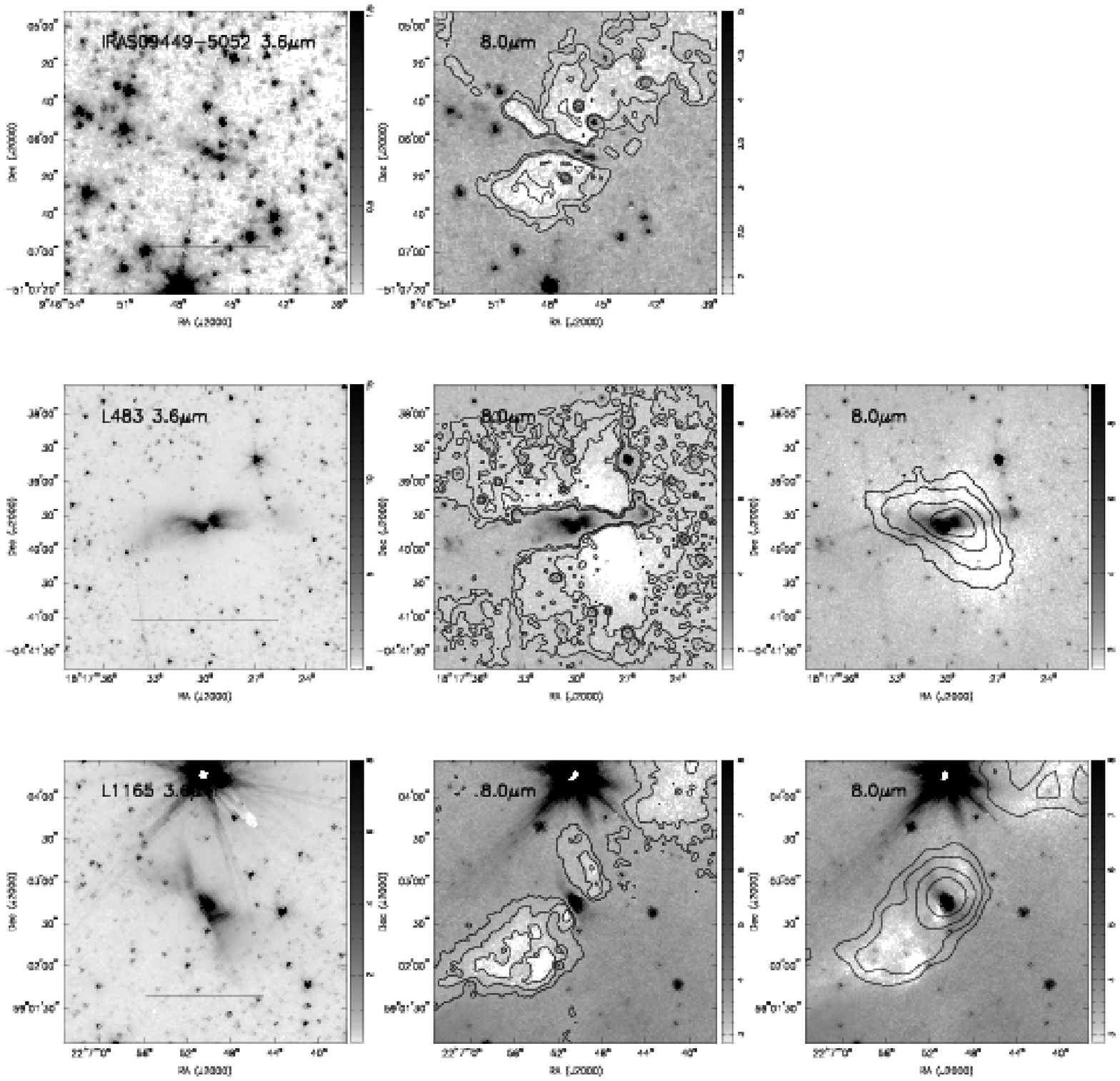}}
\centerline{Fig. 5 ---}

\begin{figure}
\begin{center}
\includegraphics[scale=0.5, angle=-90]{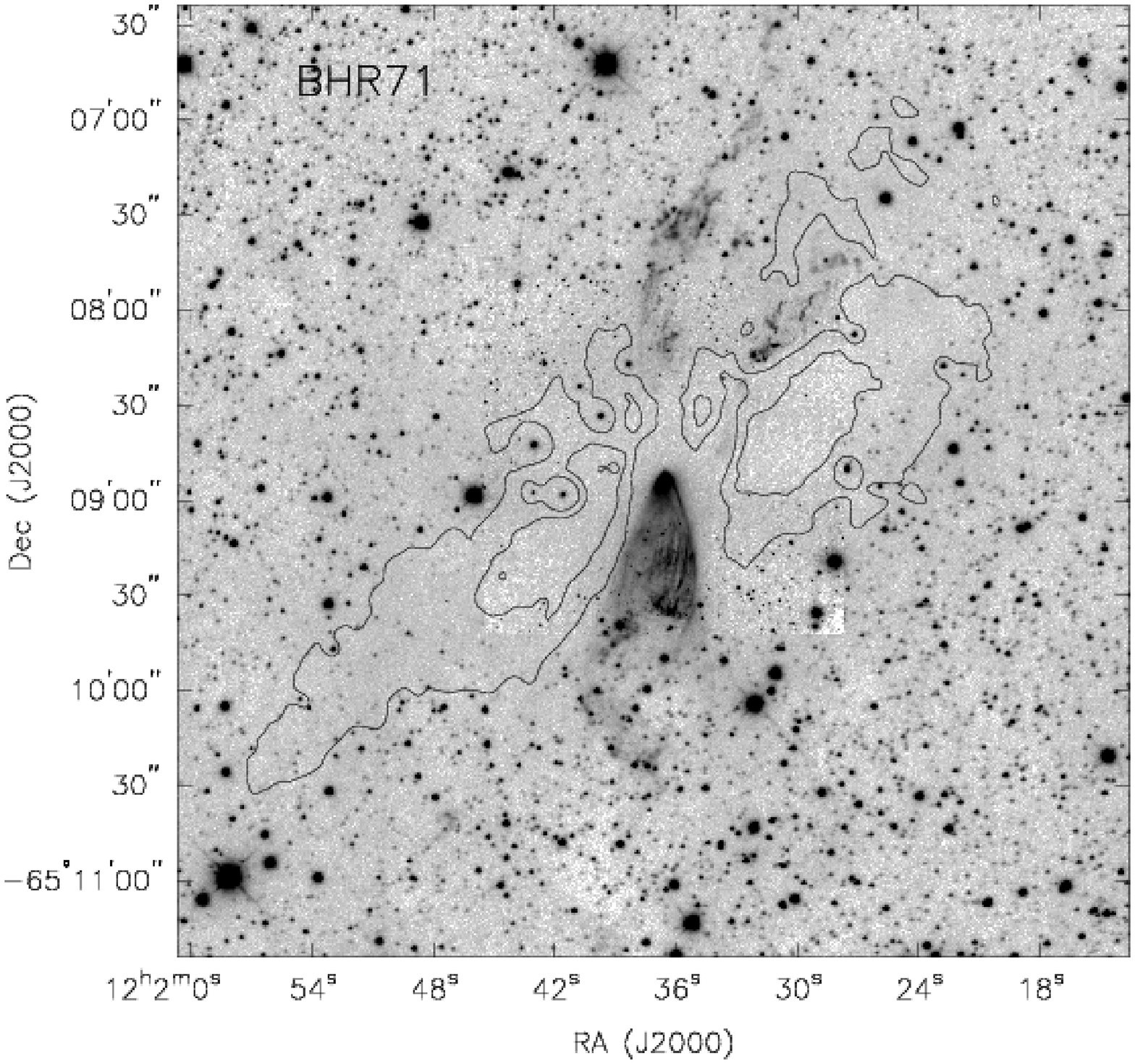}
\includegraphics[scale=0.5, angle=-90]{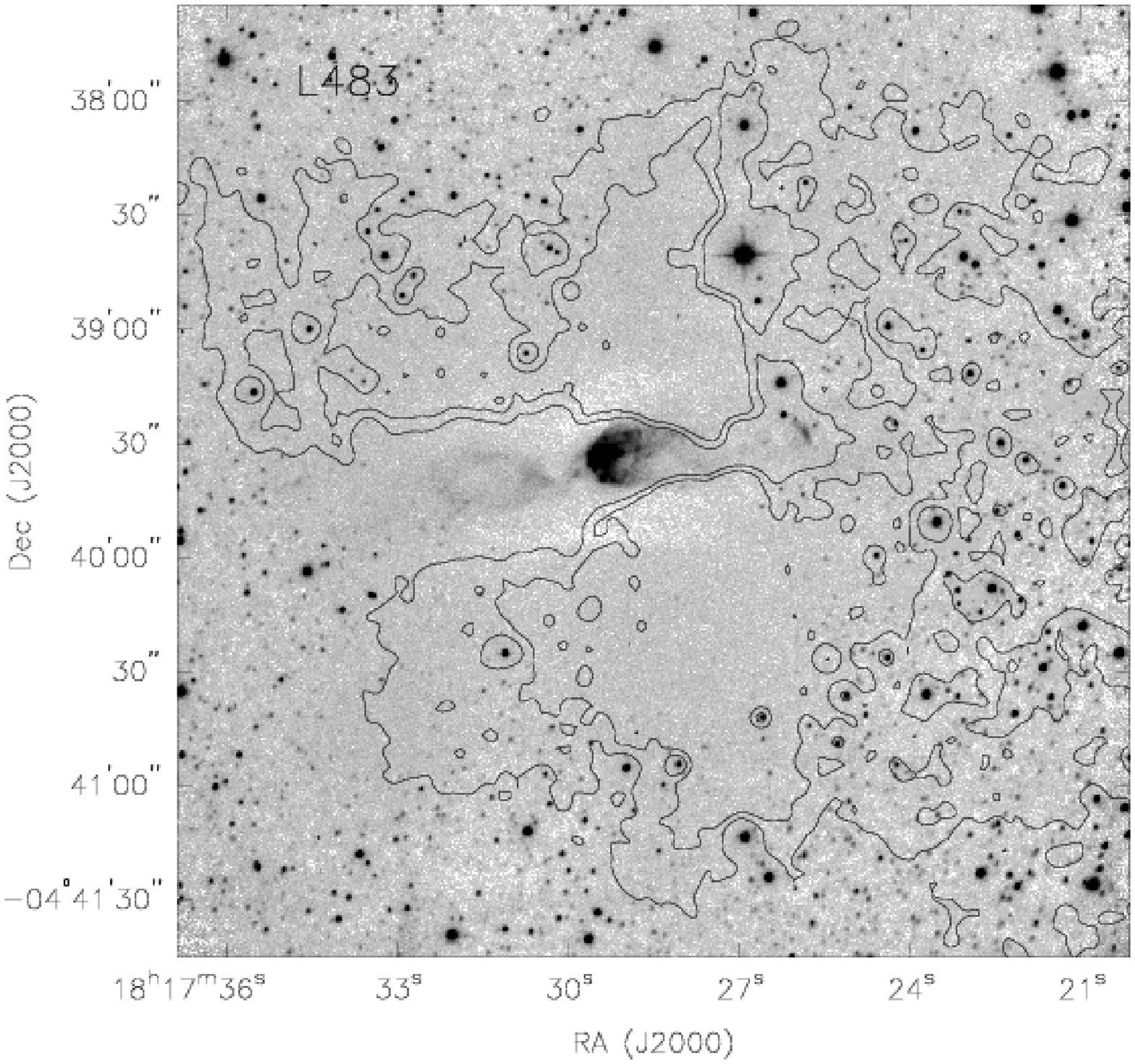}
\end{center}
\caption{Ks-band images of BHR71 and L483 with the two highest optical depth contours from Figures 1 and 5 overlaid.
As detailed in the text and Appendix, a number of background stars in the near-IR can be detected through the dense
envelope enabling the correction of the 8$\mu$m optical depth maps. The BHR71 image was taken with ISPI and 
the smaller PANIC image is inserted at the center. The L483 data was taken with TIFKAM at MDM observatory.   }
\label{nirimages}
\end{figure}

\begin{figure}
\begin{center}
\includegraphics[scale=0.5]{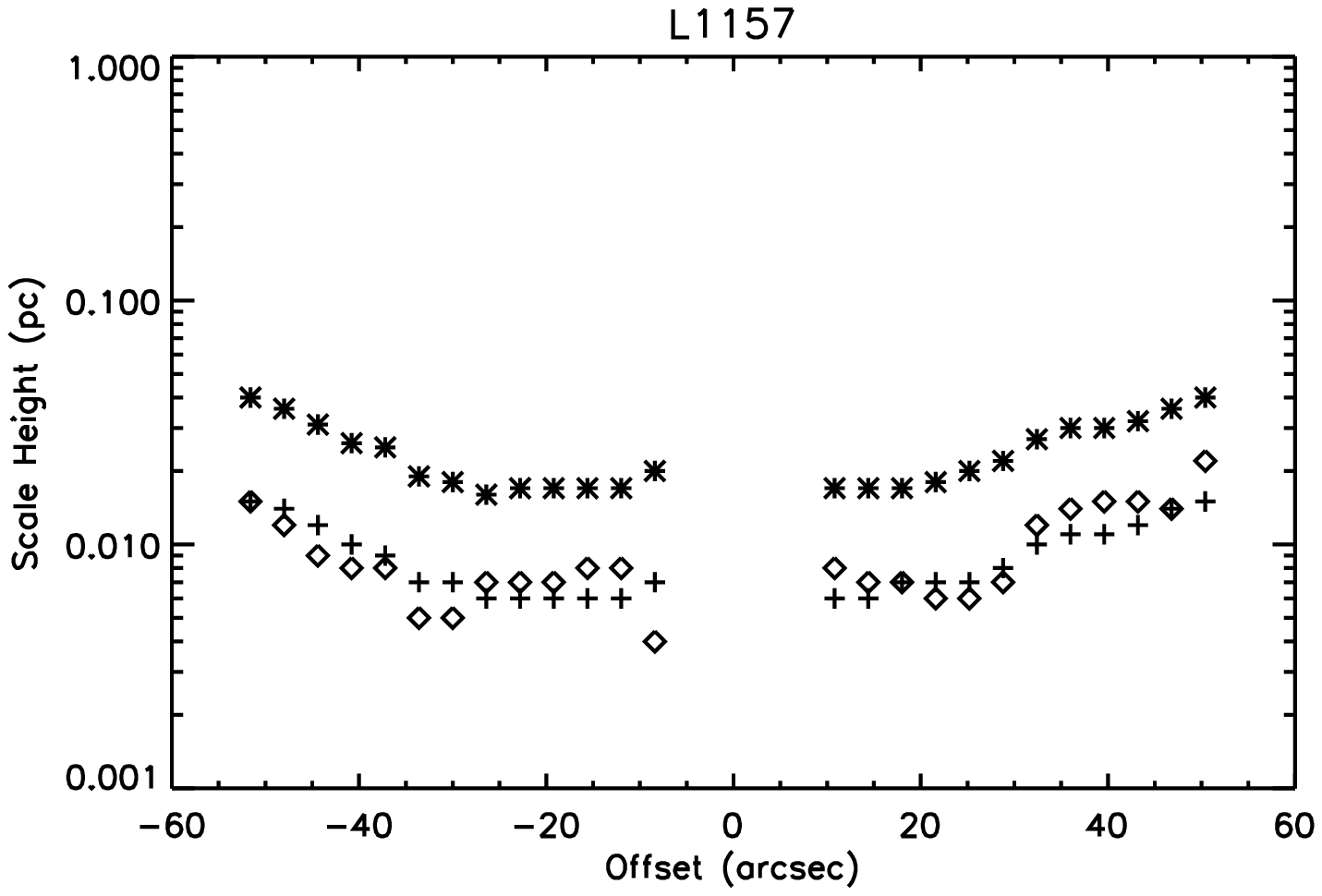}
\includegraphics[scale=0.5]{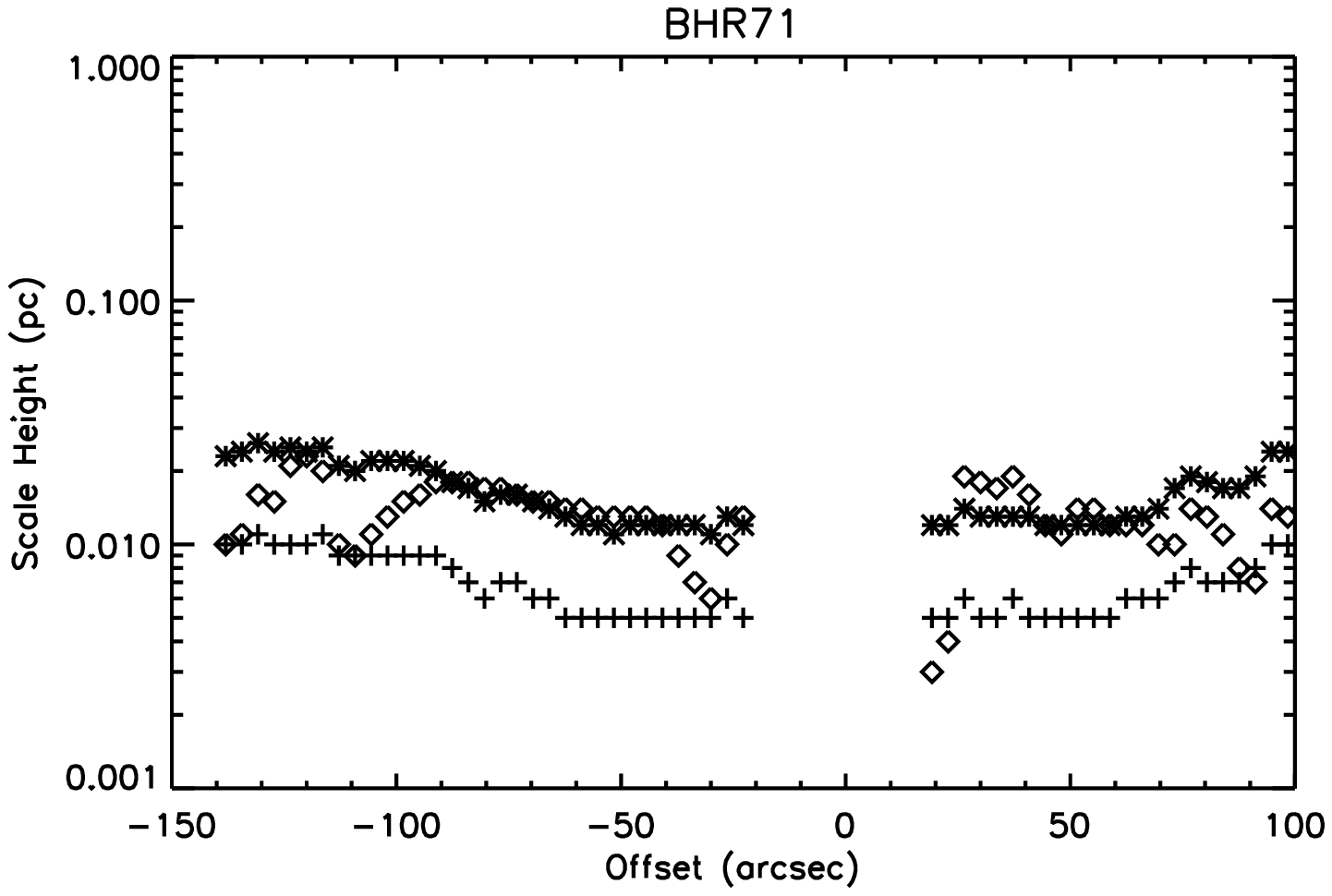}
\includegraphics[scale=0.5]{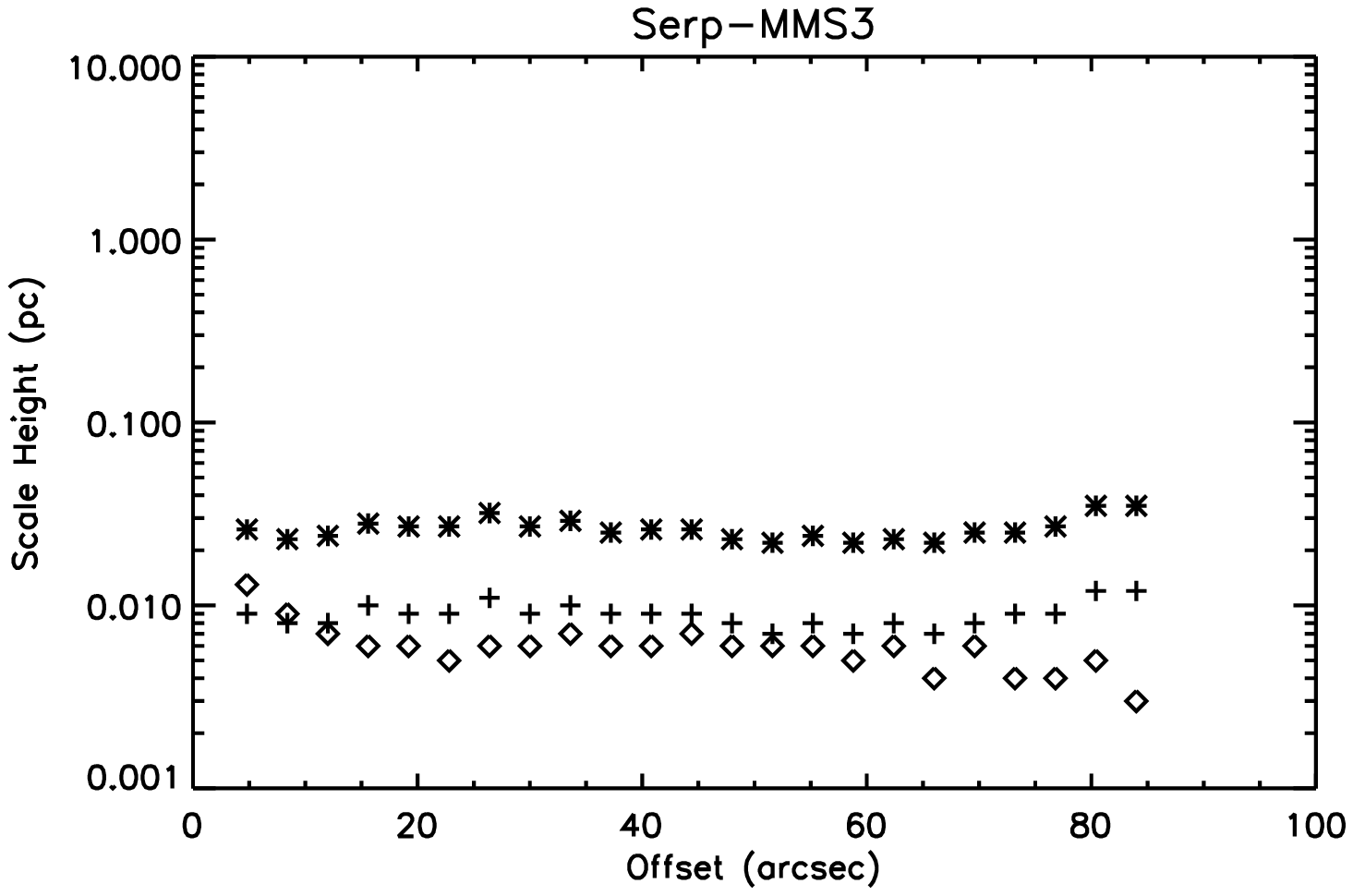}
\includegraphics[scale=0.5]{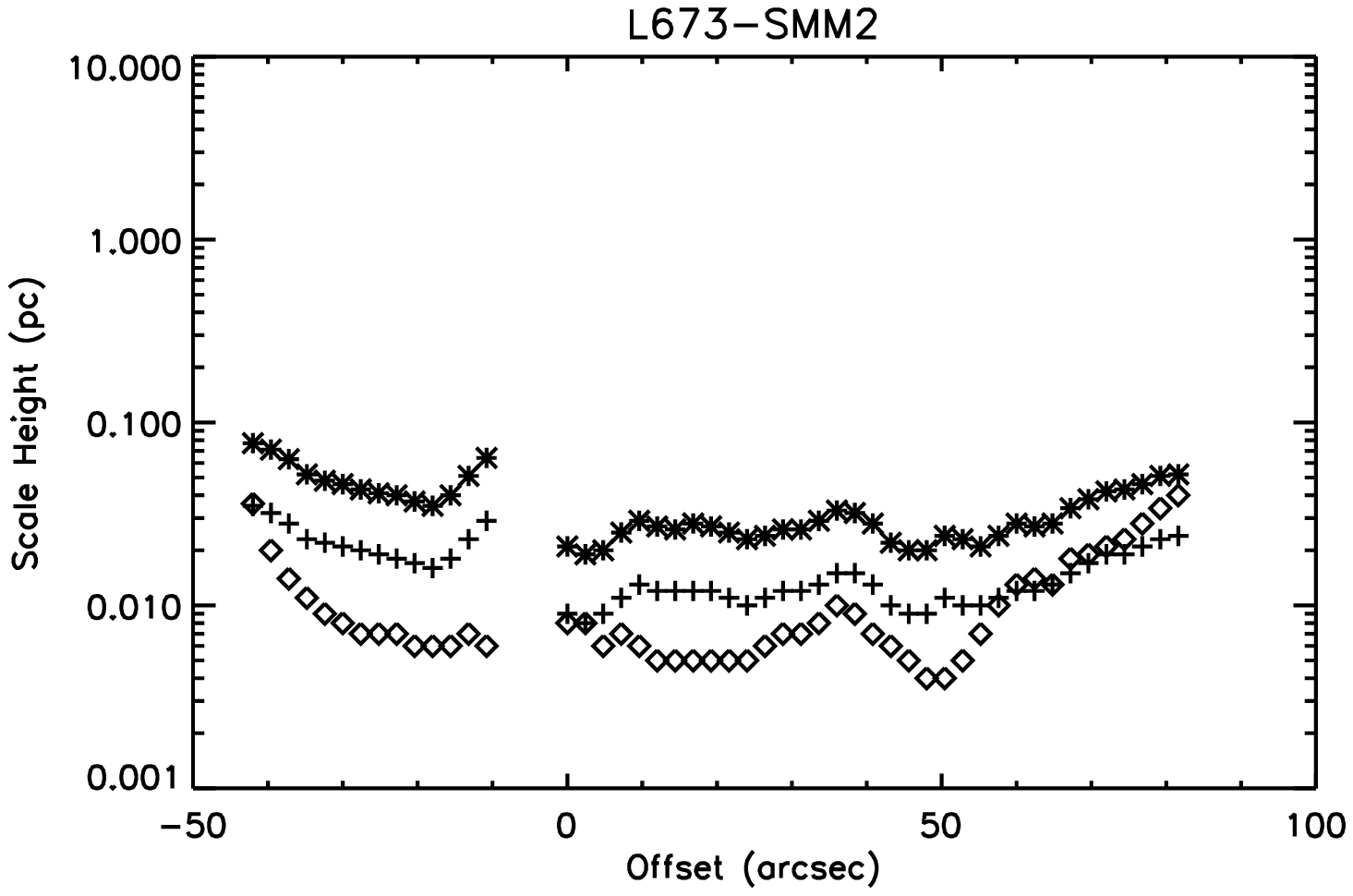}
\end{center}
\caption{Plots of scale height versus radius for flattened envelopes. The diamonds are the measured 
scale heights determined by fitting Gaussians to the data, the plus signs are the predicted
scale height of an infinite filament with peak surface density computed from the peak optical depth,
asterisks are the computed scale heights for a sheet. In the case of filaments and sheets, the scale height computed in \S3 does not
correspond to the scale height of a Gaussian, to correct for this we integrated through a filament and fit
a Gaussian to the surface density profile. The Gaussian scale height is always $\sim$1.5 $\times$ H.  }
\label{vertical}
\end{figure}

\begin{figure}
\begin{center}
\includegraphics[scale=0.75, angle=-90]{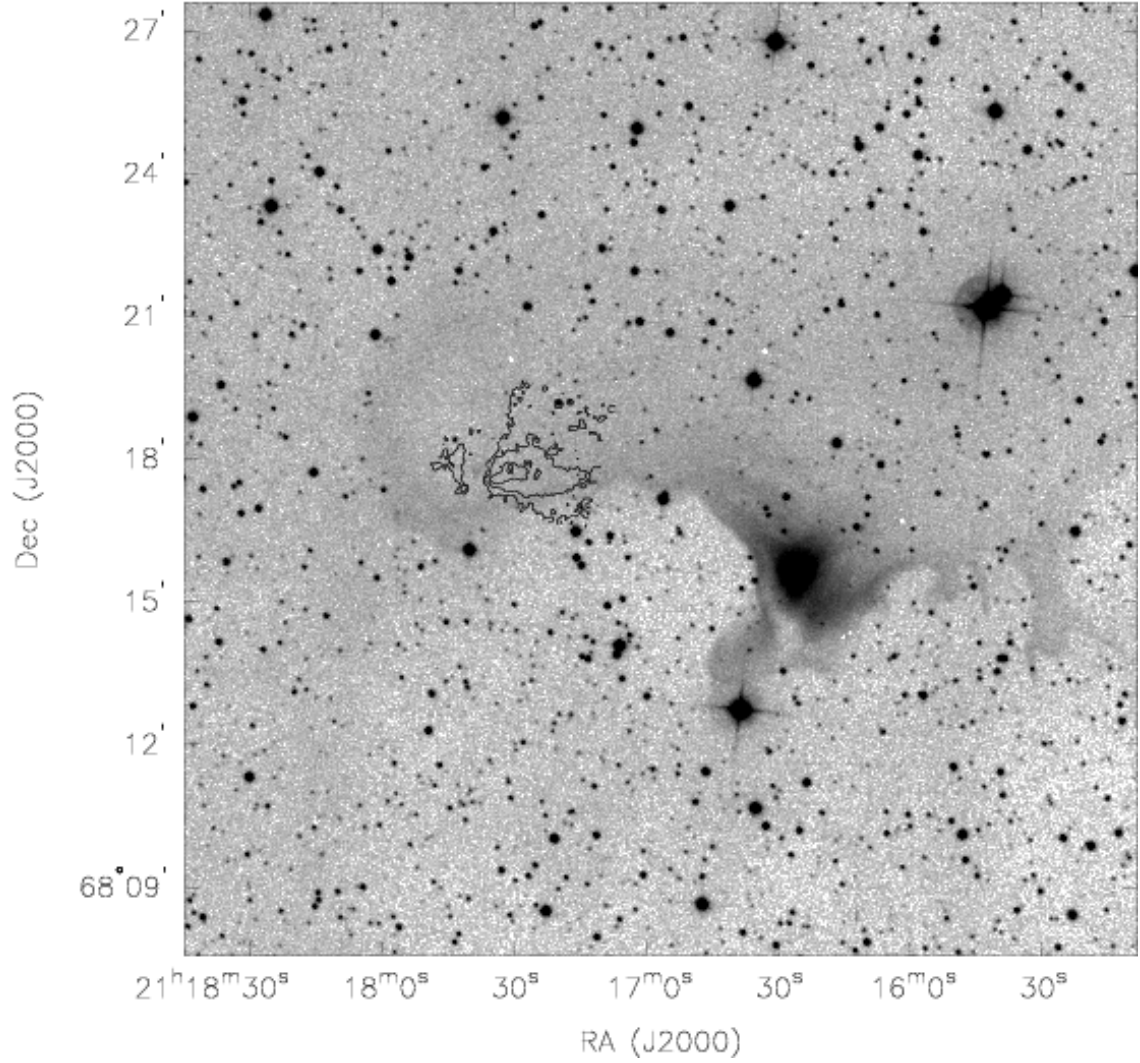}

\end{center}
\caption{DSS2 red image of CB230 region. The overlaid contours are same optical depth contours as
shown in Figure \ref{oneside} for CB230. The protostar in CB230 has formed on the far eastern side 
of the cloud while on the western side there is an optical reflection nebula possibly
from another recently formed star.}
\label{CB230}
\end{figure}

\begin{figure}
\begin{center}
\includegraphics[scale=0.7, angle=-90]{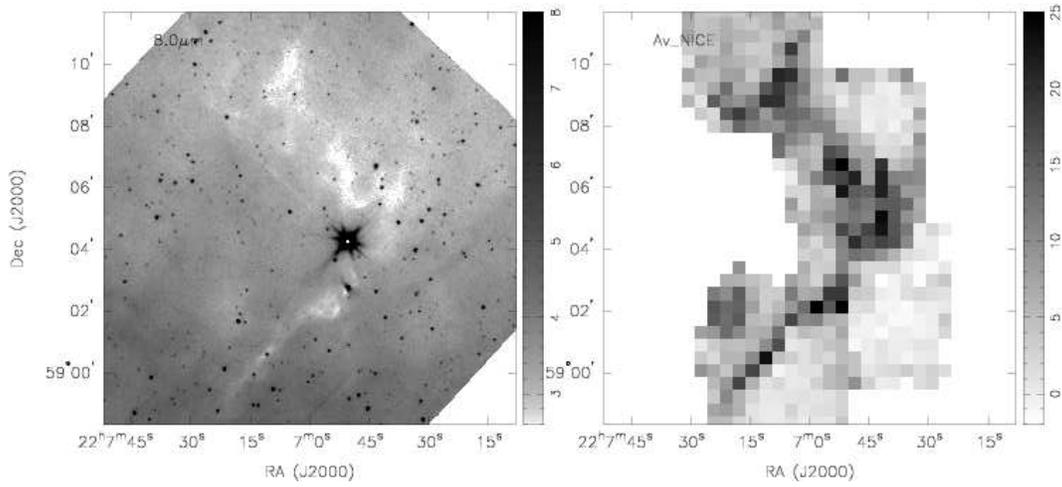}

\end{center}
\caption{Left: 8$\mu$m image of L1165 region surrounding the protostar(s). The large scale cloud structure
shows a highly filamentary structure with a 90$^{\circ}$ bend near the protostar(s). Right: Extinction
map of L1165 region constructed from H and Ks-band imaging taken with TIFKAM, units are in A$_V$. Despite the lower resolution
of the map, the large scale structure is clearly well matched between the two techniques of extinction
measurement. The large angular coverage of the dark cloud enabled good calibration of the 8$\mu$m 
optical depths using the near-IR extinction measurements.}
\label{L1165}
\end{figure}

\begin{figure}
\begin{center}
\includegraphics[scale=0.5]{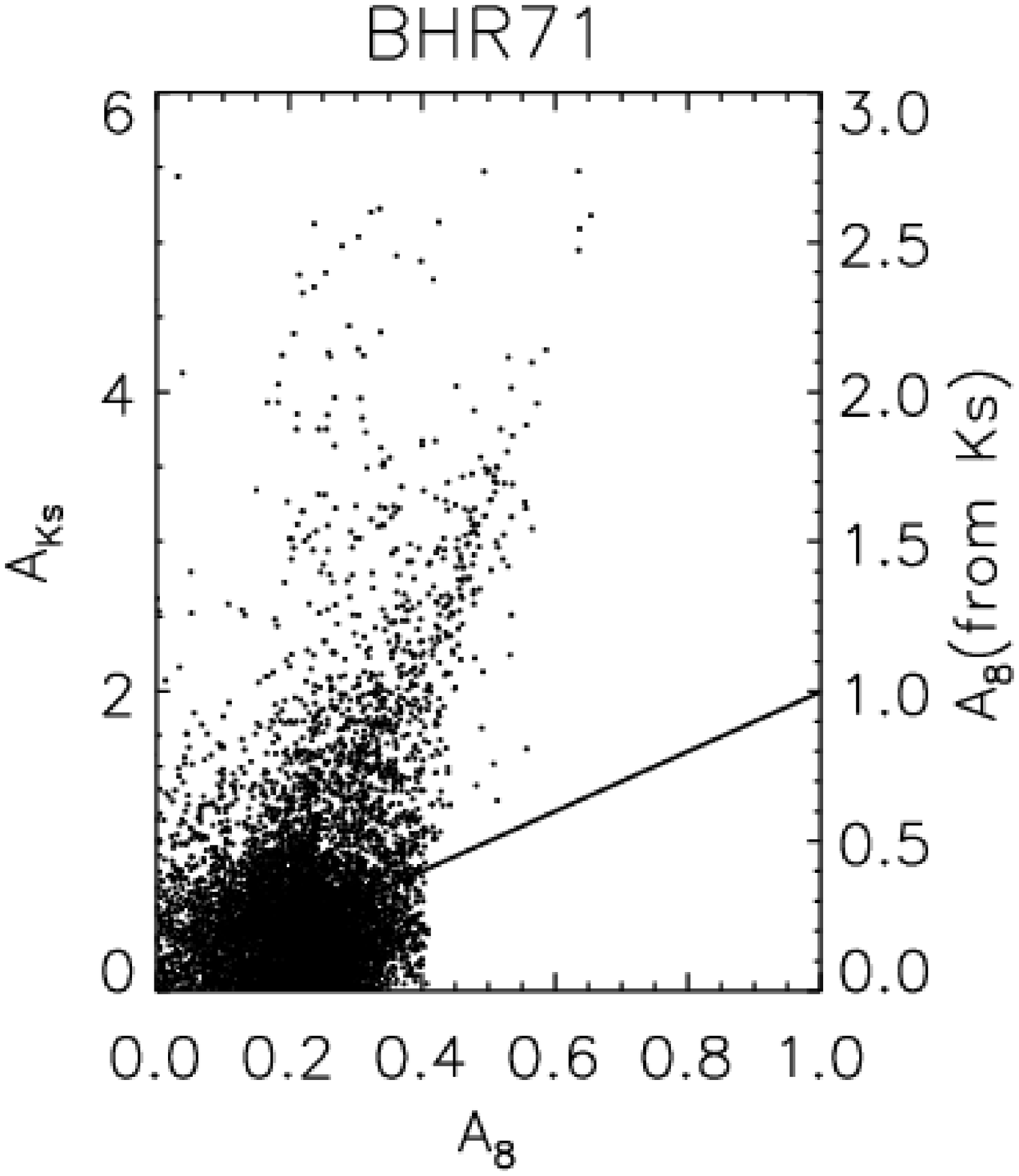}
\includegraphics[scale=0.5]{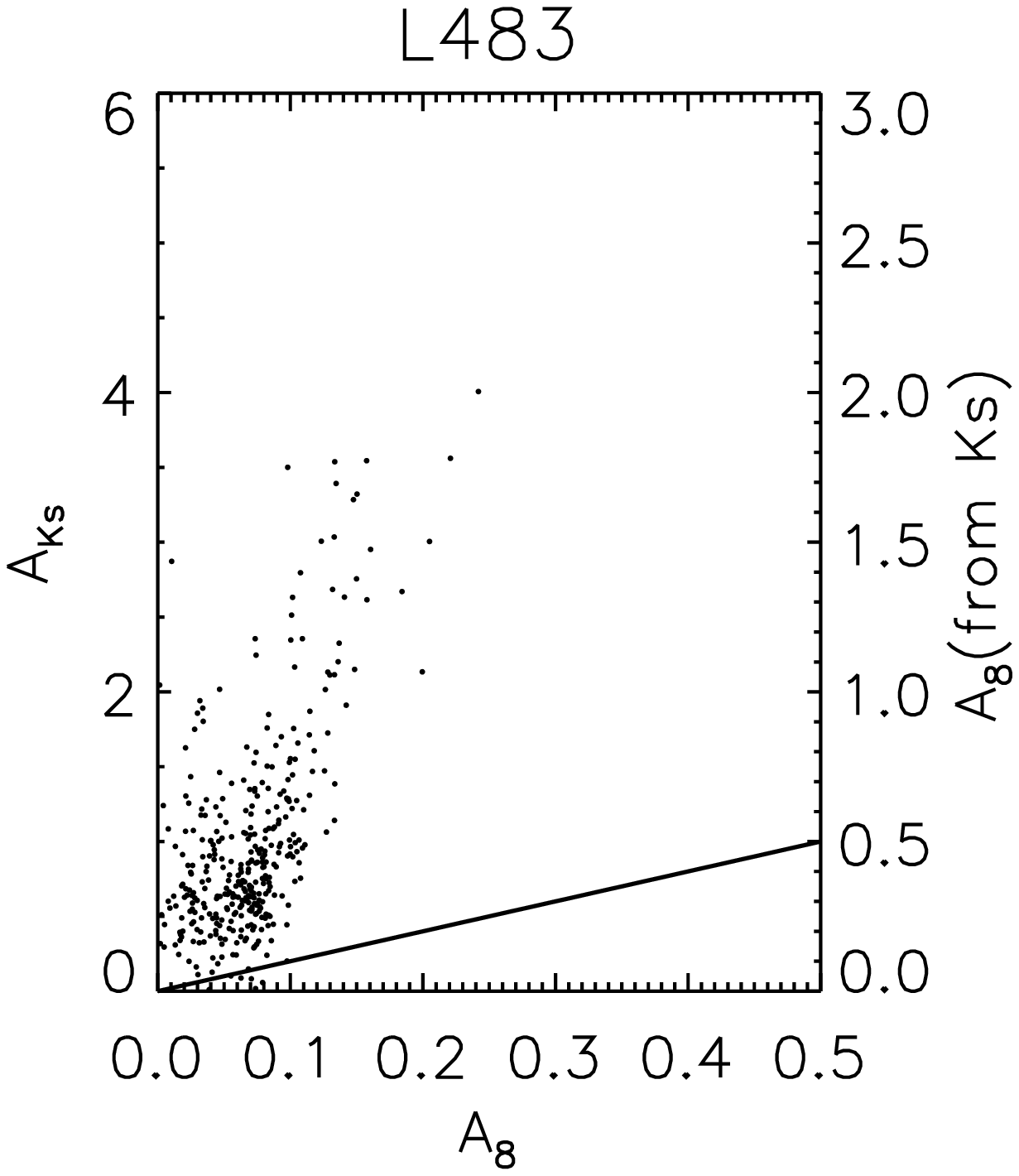}
\end{center}
\caption{Uncorrected comparison of extinction from NICE method and 8$\mu$m maps for BHR71
(Left) and L483 (Right). Solid line represents the predicted relationship between A$_{Ks}$ and A$_{8}$.}
\label{taucompnocorr}
\end{figure}

\begin{figure}
\begin{center}
\includegraphics[scale=0.5]{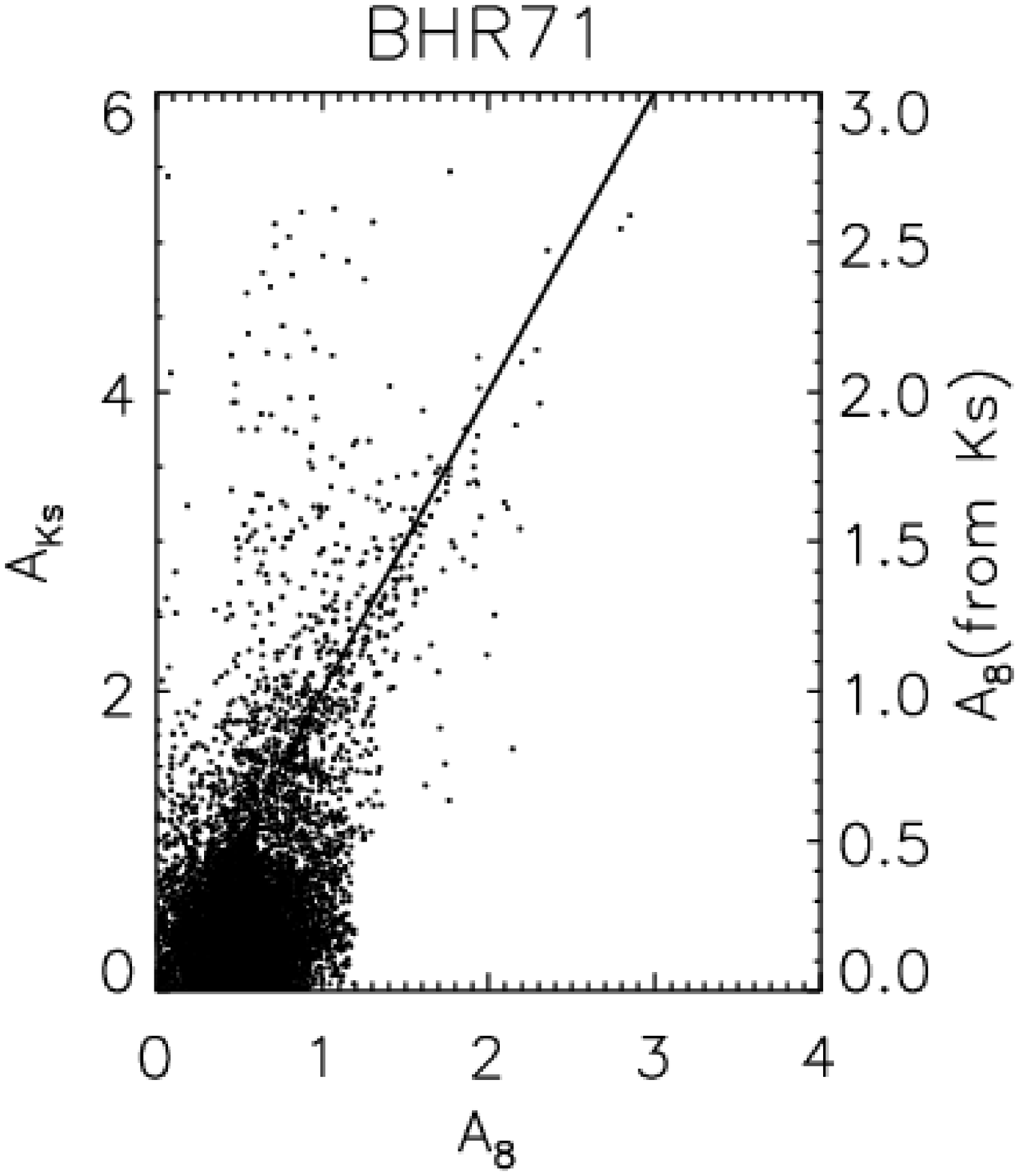}
\includegraphics[scale=0.5]{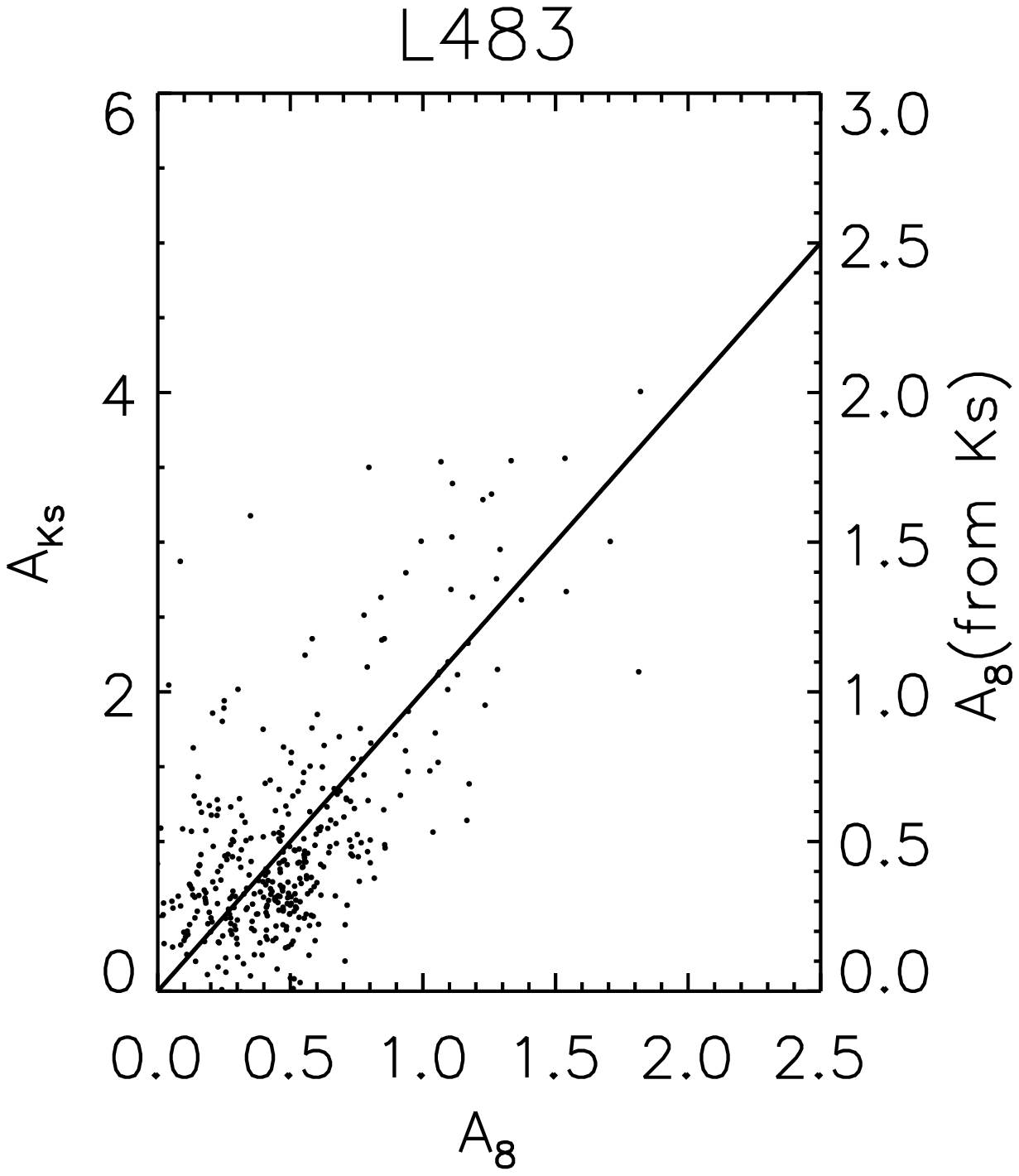}
\end{center}
\caption{ Comparison of extinction from NICE method and \textit{corrected} 8$\mu$m maps for BHR71
(Left) and L483 (Right). The points with high A$_{Ks}$ and low A$_{8}$  for BHR71 are likely due to a bright 8$\mu$m star
at the location of the near-IR measurement or scattered light and outflow emission at 8$\mu$m coincident with the
near-IR measurement.}
\label{taucompcorr}
\end{figure}
\end{small}

\clearpage
\input{tab1}
\input{tab2}
\input{tab3}
\end{document}

%% file: tab1.tex
\begin{deluxetable}{llllclc}

\tablewidth{0pt}

\tabletypesize{\scriptsize}
\tablecaption{Spitzer IRAC Observations}
\tablehead{
  \colhead{Designation}  &  \colhead{RA}      & \colhead{Dec.}      & \colhead{Date(s) Obs.} & \colhead{Int. Time} & \colhead{AORKEY/Program} & \colhead{Near-IR data}\\
                         &  \colhead{(J2000)} & \colhead{(J2000)}            &                        &            (s)             &         &   \colhead{(Obs./Inst.)} \\
}
\startdata
L1448 IRS2      & 03 25 22.5 & $+$30 45 10.5 &   2005-02-25 & 900 & 12250624/P03557 & \\
Perseus 5       & 03 29 51.6 & $+$31 39 04    &  2005-01-29 & 900 & 12249344/P03557 & \\
IRAS 03282$+$3035 & 03 31 21.0 & $+$30 45 28   &   2006-09-28 & 900 & 18326016/P30516 & \\
HH211-mm        & 03 43 56.8   & +32 00 52   &  2004-08-09 & 24 & 5790976 & \\
IRAM 04191      & 04 21 56.9 & $+$15 29 46.1 &  2005-09-18 & 480 & 14617856, 14618112 & \\
L1521F          & 04 28 39   & $+$26 51 35   &  2006-03-25 & 480 & 14605824, 14605568 & \\
L1527           & 04 35 53.9 & $+$26 03 09.7 &  2004-03-07 & 90 & 3963648 & \\
IRAS 05295$+$1247 & 05 32 19.4 & $+$12 49 41   &   2006-10-26 & 900 & 18325248/P30516 &\\
HH270 VLA1      & 05 51 34.5 & $+$02 56 48   &  2005-03-25 & 360 & 10737920 &\\
IRAS 09449-5052 & 09 46 46.5 & -51 06 07   & 2004-04-29 & 48 & 5105152 & CTIO/ISPI\\
BHR71           & 12 01 37.1 & -65 08 54   &    2004-06-10 & 150 & 5107200 & Magellan/PANIC, CTIO/ISPI\\
IRAS 16253-2429 & 16 28 22.2 & -24 36 31   &  2004-03-29 & 48 &  5762816, 5771264 &\\
L483            & 18 17 35   & -04 39 48   &  2004-09-02,03 & 48 & 5149184,5149696 & MDM/TIFKAM\\
Serp-MMS3\tablenotemark{\dagger} & 18 29 09.1 & +00 31 28.6 & 2004-04-05 & 48 & 5710848, 5712384 & \\ 
HH108           & 18 35 44.2 & -00 33 15   &  2007-05-17 & Med. Scan &14510336 & CTIO/ISPI\\
L723            & 19 17 53.2 & $+$19 12 16.6 &  2006-09-27 & 900 & 18326528/P30516 & MDM/TIFKAM\\
L673-SMM2            & 19 20 26.3 & +11 20 04   &  2004-04-30,22 & 48 & 5152256, 5151744 & \\   
L1157           & 20 39 06.2 & $+$68 02 17.3 & 2006-08-13 & 900 & 18324224/P30516 & MDM/TIFKAM\\
L1152           & 20 35 46.5 & $+$67 53 04.2 &  2004-07-23,28 & 360 & 11390976, 11399424 & MDM/TIFKAM\\
CB230           & 21 17 38.7 & $+$68 17 32.9 &  2004-11-28 & 60  & 12548864\\
L1165           & 22 06 51.0 & $+$59 02 43.5 & 2004-07-03 & 48  & 5165056 & MDM/TIFKAM\\
CB244           & 23 25 46.5 &  +74 17 39    & 2003-12-23 & 150 & 4928256 & MDM/TIFKAM\\
\enddata
\tablecomments{} 
\tablenotetext{\dagger}{Source is identified as MMS3 in \citet{dab2006}.}
\end{deluxetable}

%% file: tab2.tex
\begin{deluxetable}{lcccccclllllc}
\rotate
\tablewidth{0pt}

\tabletypesize{\scriptsize}
\tablecaption{Source Properties}
\tablehead{
  \colhead{Designation}  & \colhead{Distance}  & \colhead{Mass} & \colhead{Mass} & \colhead{Mass} & \colhead{Mass$_{obs}$} &\colhead{L$_{bol}$}  & \colhead{$\sigma\tau$} & \colhead{$\tau_{max}$ } & \colhead{$I_{bg}^{\prime}$} & \colhead{$I_{fg}$} & \colhead{Classification}& \colhead{References}\\
                         & \colhead{(pc)}      &    \colhead{($M_{\sun}$)}&    \colhead{($M_{\sun}$)}&    \colhead{($M_{\sun}$)}   &     \colhead{($M_{\sun}$)} &     \colhead{($L_{\sun}$)}  & & & (MJy/sr)  & (MJy/sr)  &   &\colhead{(M$_{obs}$, L$_{bol}$)}  \\
                         &                     &  \colhead{(r$<$0.15pc)} &  \colhead{(r$<$0.1pc)} &  \colhead{(r$<$0.05pc)}
}
\startdata
L1448 IRS2      & 300 & 23.0 & 9.8 & 2.1 &  0.86\tablenotemark{*} & 8.4    & 0.15 & 2.35 & 3.98 & 3.44 & Flattened & 13,1\\
Perseus 5       &  300 & 19.2 & 9.8 & 3.4 & 1.24\tablenotemark{*} & 0.46   & 0.15 & 2.3 & 4.04 & 3.56  & One-sided & 3,2\\
IRAS 03282+3035 & 300 & 25.7 & 14.2 & 4.0 & 2.2\tablenotemark{*}  & 1.2    & 0.3 & 2.1 & 2.55 & 2.26   & Irregular & 4,2\\
HH211           & 300 & 5.5 & 3.4 & 1.8 &   1.5\tablenotemark{*}  & 3.02   & 0.15 & 1.7  & 10.1 & 8.57 & Irregular & 3,3\\
IRAM 04191      & 140 & 9.7 & 6.1 & 1.0 & ...    & 0.3    & 0.22  & 0.9 & 4.09 & 3.77 & Spheroidal & ..,7\\
L1521F          & 140 & 5.9 & 4.8 & 2.3 & ...    & 0.03   & 0.35  & 1.2 & 4.65 & 4.35 & Spheroidal & ..,8\\
L1527           & 140 & 1.9 & 1.5 & 0.84 & 2.4\tablenotemark{\dagger}   & 2.0    & 0.2  & 1.5 & 4.37 & 4.0   & One-sided & 5,9\\
IRAS 05295+1247 & 400 & 2.9 & 2.4 & 1.5 & ...    & 12.5    & 0.05 & 3.3 & 5.2 & 4.02   & Irregular &..,1\\
HH270 VLA1      & 420 &  7.8 & 5.7 & 1.9  & ...  & 7.0    & 0.12 & 2.4 & 4.2 & 3.72   & One-sided & ..,1\\
IRAS 09449-5052 & 300? & 4.7 & 3.6 & 1.8  & 3.7\tablenotemark{\dagger}  & 3.1    & 0.25 & 1.25 & 2.46 & 2.1  & Irregular & 6,6\\
BHR71           & 200 & 24.5 & 13.8 & 4.6 & 2.2\tablenotemark{\dagger}  & 9.0    & 0.1 & 2.8 & 4.36 & 2.3    & Flattened & 6,6\\
IRAS 16253-2429 & 125 & 5.63 & 2.9 & 0.8 & 0.98\tablenotemark{*}  & 0.25   & 0.1  & 1.4 & 7.0 & 4.52   & Spheroidal & 10,2\\
L483            & 200 & 21.5 & 13.2 & 3.5 & 1.8\tablenotemark{*}  & 11.5   & 0.25 & 1.75 & 2.87 & 2.35 & Irregular &  4,1\\
Serp-MMS3& 225& 3.1  & 1.3  & 0.3 & 2.2\tablenotemark{*}  &  1.6  & 0.1  & 2.20 & 3.93 & 1.48 & Flattened &  2, 15 \\
HH108           & 300 &      &      &     &  4.5, 3.6\tablenotemark{*} & $\sim$8.0, 1.0    &       &     &      &      & Flattened &  14\\
L723            & 300 & 4.0 & 2.1 & 0.6 & 1.6\tablenotemark{*}    & 4.6   & 0.03 & 3.1 & 3.22 & 2.55  & Flattened & 4,1\\
L673-SMM2            & 300 & 5.2 & 3.2 & 1.0 & 0.35\tablenotemark{*}   &  2.8 & 0.1  & 3.6 & 9.39 & 3.33  & Flattened & 12 \\
L1157           &  250 & 12.0 & 6.0 & 0.6 & 2.2\tablenotemark{*} & 3.0   & 0.1 & 2.5 & 0.71 & 0.45   & Flattened   & 4,1\\
L1152           & 250 & 14.1 & 7.8 & 2.4  & 12.0\tablenotemark{\dagger} & 1.3   & 0.15 & 1.85 & 0.32 & 0.18 & Binary Core & 5,1\\
CB230           & 300 & 7.2 & 4.1 & 1.1   & 1.1\tablenotemark{*}  & 7.2   & 0.23 & 1.9 & 1.79 & 1.17  & One-sided & 11,1\\
L1165           & 300 & 6.9 & 3.8 & 1.1 & 0.32\tablenotemark{*}   & 28     & 0.1 & 3.0 & 3.76 & 2.6    & Irregular & 12,12\\
CB244           & 250 &  9.7 & 7.8 & 4.1 & ...   & ...     & 0.2 & 1.75 & 0.59 & 0.35  & Binary Core & ..,..\\
\enddata
\tablecomments{ For HH108 the quoted values are given for IRS1 and IRS2 respectively.
References: (1) This work, (2) \citet{enoch2009}, (3)\citet{enoch2006},
 (4) \citet{shirley2000}, (5) \citet{benson1989}, (6) \citet{bhr1995}, (7) \citet{dunham2006}
 (8) \citet{terebey2009}, (9) \citet{tobin2008}, (10) \citet{young2006}, (11) \citet{kauffmann2008}
 (12) \citet{visser2002}, (13) \citet{olinger1999}, (14) \citet{chini2001}, (15) \citet{enoch2007}.} 
\tablenotetext{*}{Mass is computed with sub/millimeter bolometer data assuming an isothermal temperature.}
\tablenotetext{\dagger}{Mass is computed from ammonia maps assuming an abundance relative to H$_2$ and excitation temperature.}
\end{deluxetable}

%% file: tab3.tex
\begin{deluxetable}{lcccc}
\rotate
\tablewidth{0pt}

\tabletypesize{\scriptsize}
\tablecaption{Moment of Inertia Ratios}
\tablehead{
  \colhead{Object}  &  \colhead{Radius} & \colhead{$I_{\perp}$/$I_{\parallel}$}  &    \colhead{$I_{\perp,l}$/$I_{\perp,r}$} &    \colhead{$I_{\parallel,l}$/$I_{\parallel,r}$}  \\
                     & \colhead {(pc)}\\
}
\startdata
Perseus 5 & 0.05 & 1.2 & 1.2 & 1.0\\
L1448 IRS2 & 0.05 & 0.6 & 1.0 & 1.7\\
IRAS 03282+3035 & 0.05 & 1.2 & 1.0 & 0.8\\
HH211 & 0.05 & 2.2 & 0.2 & 0.2\\
L1448-IRS2 & 0.05 & 1.7 & 0.6 & 1.0\\
IRAM 04191 & 0.03 & 1.5 & 2.5 & 1.2\\
L1521F & 0.04 & 1.3 & 0.9 & 1.0\\
L1527 & 0.05 & 3.2 & 0.1 & 0.1\\
IRAS 05295+1247 & 0.05 & 1.3 & 0.5 & 0.5\\
HH270 VLA1 & 0.10 & 0.6 & 1.4 & 2.6\\
IRAS 09449-5052 & 0.05 & 1.7 & 0.8 & 1.0\\
BHR71 & 0.075 & 1.6 & 1.1 & 0.8\\
IRAS 16253-2429 & 0.05 & 1.1 & 0.9 & 1.0\\
L483 & 0.05 & 1.4 & 1.0 & 0.9\\
Serp-MMS3 &  0.05 & 1.0 & 1.9 & 1.5\\
L723 & 0.05 & 1.2 & 0.8 & 0.9\\
L673-SMM2 & 0.05 & 0.9 & 0.1 & 0.4\\
L1157 & 0.05 & 3.0 & 0.9 & 0.8\\
L1152 & 0.05 & 0.8 & 1.4 & 1.5\\
CB230 & 0.05 & 3.4 & 0.3 & 0.6\\
L1165 & 0.05 & 2.3 & 0.5 & 0.8\\
\enddata
\tablecomments{$I_{\perp}$ measures material that is located away from the outflow/rotation axis along the abscissa and $I_{\parallel}$
 measures material located along the ordinate axis. CB244 does not appear in this table because 
most extinction in this system is associated with the neighboring starless core, not the protostar.
 } 
\end{deluxetable}